\documentclass[a4paper, notoc]{JHEP3}
\usepackage{amsfonts}
\usepackage{amssymb}
\usepackage{comment}
\usepackage{epsfig}
\usepackage{marginnote}
\usepackage{graphicx}
\usepackage{amsmath}
\usepackage{enumerate}
\usepackage{cite}

\newcommand{\be}{\begin{equation}}
\newcommand{\ee}{\end{equation}}
\newcommand{\bea}{\begin{eqnarray}}
\newcommand{\eea}{\end{eqnarray}}
\newcommand{\mbb}{\mathbb}

\newcommand{\mc}{\mathcal}

\newcommand{\vo}{{\cal{V}}}

\newcommand{\Zbb}{\mathbb{Z}}

\newcommand{\G}{\Gamma}

\newcommand{\ti}{r}
\newcommand{\ri}{t}

\title{Moduli Stabilisation for Chiral Global Models}

\author{Michele Cicoli${}^{1,2}$, Christoph Mayrhofer${}^3$, Roberto Valandro${}^4$ \\

$^1$ Deutsches Elektronen-Synchrotron DESY, Hamburg, Germany.\\
Email: \email{michele.cicoli@desy.de} \\
$^2$ Abdus Salam ICTP, Strada Costiera 11, Trieste 34014, Italy. \\
Email: \email{mcicoli@ictp.it} \\
$^3$ Institut f\"ur Theoretische Physik, Universit\"at Heidelberg, Heidelberg, Germany.\\
Email: \email{c.mayrhofer@thphys.uni-heidelberg.de} \\
$^4$ II. Institut f\"ur Theoretische Physik, Universit\"at Hamburg, Hamburg, Germany. \\
Email: \email{roberto.valandro@desy.de}}

\abstract{We combine moduli stabilisation and (chiral) model building in a fully consistent global set-up in Type IIB/F-theory.
We consider compactifications on Calabi-Yau orientifolds which admit an explicit description in terms of toric geometry.
We build globally consistent compactifications with tadpole and Freed-Witten anomaly cancellation
by choosing appropriate brane set-ups and world-volume
fluxes which also give rise to $SU(5)$- or MSSM-like chiral models.
We fix all the K\"ahler moduli within the K\"ahler cone
and the regime of validity of the 4D effective field theory.
This is achieved in a way compatible with the local presence of chirality. The hidden sector generating the non-perturbative effects is placed on a del~Pezzo divisor that does not have any chiral intersection with any other brane.
In general, the vanishing D-term condition implies the shrinking
of the rigid divisor supporting the visible sector.
However, we avoid this problem by generating $r<n$ D-term conditions on a set of $n$ intersecting divisors.
The remaining $(n-r)$ flat directions are fixed by perturbative corrections to the K\"ahler potential.
We illustrate our general claims in an explicit example. We consider a K3-fibred Calabi-Yau with four K\"ahler moduli, that is a hypersurface in a toric ambient space and admits a `simple' F-theory up-lift.
We present explicit choices of brane set-ups and fluxes which
lead to three different phenomenological scenarios:
the first with GUT-scale strings and TeV-scale SUSY by fine-tuning the background fluxes; the second with an exponentially large value of the volume and
TeV-scale SUSY without fine-tuning the background fluxes;
and the third with a very anisotropic configuration that leads
to TeV-scale strings and two micron-sized extra dimensions.
The K3 fibration structure of the Calabi-Yau three-fold
is also particularly suitable for cosmological purposes.}

\preprint{DESY-11-179\\ ZMP-HH/11-15}

\begin{document}

\tableofcontents

\section{Introduction}

Two longstanding problems of Calabi-Yau compactifications, whose
solution is crucial for string theory to make contact with particle phenomenology and cosmology, are moduli stabilisation
and the derivation of GUT- or MSSM-like constructions.

The discovery of D-branes in 1995 \cite{Polchinski:1995mt} opened the possibility to solve both of these problems at the same time within Type II theories. In fact, D-branes, on one hand, provide non-Abelian gauge symmetries and chiral
matter, whereas, on the other hand, are needed to turn
on background fluxes \cite{deWit:1986xg,Maldacena:2000mw}. These fluxes play a key-r\^ole to
fix most of the moduli in Type IIB theories due to their mild back-reaction on the Calabi-Yau geometry even if they do not develop a potential for the K\"ahler moduli \cite{gkp,Dasgupta:1999ss} (for a review see \cite{Grana:2005jc,fluxesDoKa,hepth0610327,Denef:2008wq}).

Given that K\"ahler moduli stabilisation has to be studied globally
while model building is mainly a local issue,
the solution of these two problems has been thought to decouple. Therefore,
a lot of effort has been put on trying to solve them separately.
This trend has also been due to the intrinsic difficulty to
address these issues simultaneously. Some viable mechanisms to fix the K\"ahler moduli
are now available \cite{kklt,LVS,bhkStab}. Moreover, after a fruitful development in local model building with intersecting and magnetised D7-branes, both in perturbative Type IIB and in F-theory (see for example  \cite{Cascales:2003wn,Marchesano:2004xz,Cvetic:2005bn,Acharya:2006mx,Donagi:2008ca,BHV,Hayashi:2008ba,CMQ,Marsano:2009gv,Conlon:2009qq,CamDudPal2011
}), also global realistic models have been constructed \cite{BBGW,CKMW,Blumenhagen:2009yv,Marsano:2009wr,Grimm:2009yu,Cvetic:2010rq,Chen:2010ts,diss}.
However, it is time to try to combine the solutions of the two problems together.

The first attempts to do so \cite{BBGW,CKMW,blumenhagen} realised that moduli stabilisation
and model building are not completely decoupled,
and so they cannot be consistently studied separately. Apart
from the standard issues of having robust control over the
effective field theory and the stabilisation of the moduli within
the K\"ahler cone, there are three crucial problems to solve:
\begin{itemize}
\item[$\ast$] Tension between moduli stabilisation via non-perturbative effects
and chirality \cite{blumenhagen};
\item[$\ast$] Tension between moduli stabilisation via non-perturbative effects and the
cancellation of Freed-Witten anomalies \cite{BBGW,CKMW};
\item[$\ast$] Various divisors, among which the one supporting the visible sector,
might be forced to shrink to zero size by the D-term constraints \cite{BBGW,CKMW,CKM}.
\end{itemize}

In this paper we present some models where all these problems can be
solved simultaneously as well as obtaining very interesting phenomenological scales.
Therefore, this work sets the basis for the realisation of a consistent realistic model within the framework of Type IIB Calabi-Yau flux compactifications where issues like moduli stabilisation, supersymmetry breaking and inflation,
can be successfully combined together with issues like MSSM-like constructions with the right chiral spectrum and Yukawa couplings.

More in detail, we present Type IIB chiral models with the following features:
\begin{itemize}
\item We focus on Type IIB/F-theory string compactifications with D3/D7-branes and O3/O7-planes;

\item We give an explicit description of the compact Calabi-Yau. By means of toric geometry we derive the topology of all toric divisors and the K\"ahler cone of the three-fold;

\item We specify two explicit brane set-ups with a particular choice of world-volume fluxes that produces
an $\mc{N}=1$ chiral model (either $SU(5)$- or MSSM-like).
These set-ups pass several global consistency checks like D7-tadpole, K-theory charges and Freed-Witten anomaly cancellation. Moreover, considering all the sources for D3-charge, our constructions
leave enough space to turn on background three-form fluxes
that would fix the complex structure moduli and still satisfy the D3-tadpole cancellation condition;

\item We stabilise all the K\"ahler moduli in a way compatible with the local presence of chirality
and avoiding the possible shrinking of any divisor induced by D-terms.
All the K\"ahler moduli are fixed within the K\"ahler cone and the regime of validity
of the effective field theory reproducing a visible sector gauge coupling of the correct size;

\item In our two constructions, we present three choices of underlying parameters which
produce three different models with interesting phenomenological scales:
\begin{enumerate}
\item GUT-scale strings and TeV-scale SUSY by fine-tuning the background fluxes;

\item Intermediate scale strings and TeV-scale SUSY for natural values of the background fluxes;

\item TeV-scale strings and micron-sized extra dimensions for very anisotropic compactifications \cite{ADDstrings}.
\end{enumerate}

\item The last two examples represent the first realisation of LARGE Volume Scenarios (LVS) \cite{LVS} for globally consistent chiral models.

\item Our K3-fibred Calabi-Yau constructions are particularly promising for cosmological applications \cite{FibreInfl}.
\end{itemize}

In spite of all these successes, there are still many issues to be addressed: like
the explicit analysis of three-form background fluxes to fix the complex structure, the dilaton and D7-brane deformation moduli;
the realisation of a fluxed brane setup that produces the right
chiral spectrum and Yukawa couplings; the implementation of
the up-lift to a Minkowski vacuum and the derivation of all the details of the inflationary scenario.
We leave all these issues for future investigation. Nevertheless,
we believe this work to represent already a big step forward.

This paper is organised as follows. In section \ref{GenDisc} we first set the stage
for our analysis by briefly reviewing the main problems that one encounters when
trying to combine moduli stabilisation with local chiral D-brane models. Then, we present our general strategy to overcome all these difficulties and
obtain a viable compact chiral model with all closed string moduli stabilised.
In section \ref{TheExample} we illustrate our general claims in an
explicit example of a K3-fibred Calabi-Yau three-fold with four K\"ahler moduli
taken from the list of \cite{CKM}. We outline all the details of a GUT-like model
with two D-term conditions in section \ref{Ex1} while in section \ref{Ex2}
we describe a second model with just one D-term condition which
yields an exponentially large volume for the internal three-fold and two different sets of phenomenologically interesting scales.
Finally, we give our conclusions in section \ref{Concl}.

\section{Type IIB moduli stabilisation}
\label{GenDisc}

In this section we shall illustrate our general procedure
to fix all the closed string moduli of Type IIB flux compactifications.
We shall focus on compactifications on Calabi-Yau orientifolds $X$ compatible
with the presence of D3/D7-branes and O3/O7-planes. These
lead to a low-energy $\mc{N} = 1$ (supergravity) effective field theory, below the Kaluza-Klein scale.

\subsection{Tree-level stabilisation}

The closed string moduli of the $\mc{N}=1$ 4D supergravity, obtained as the low-energy limit
of Type IIB string theory, are given by the Kaluza-Klein reduction of the massless bosonic fields of the 10D theory.
In the Ramond-Ramond sector these include the even forms $C_{2p}$, $p=0,1,2$, while
the Neveu Schwarz-Neveu Schwarz sector involves the dilaton $\phi$, the metric $g^s_{MN}$ (in string frame)
and the antisymmetric two-form $B_2$.

The 4D closed string moduli are then given by the axio-dilaton $S=e^{-\phi }+i C_0$,
the complex structure moduli $U_{\alpha}$, $\alpha=1,...,h_-^{2,1}(X)$, and the K\"ahler moduli:
\bea
T_i&=&\tau_i+i\, \zeta_i^+,\quad\tau_i=\textrm{Vol}(D_i),\quad \zeta_i^+=\int_{D_i}C_4,\quad i=1,...,h^{1,1}_+\,, \nonumber \\
G_j&=&c_j-i S \zeta_j^-,\quad C_2=\sum_j c_j \, \hat{D}_j,\,\,\,\, B_2=\sum_j \zeta_j^- \hat{D}_j ,\quad j=1,...,h^{1,1}_-\,. \nonumber
\eea
Here $D_i$ denotes a four-cycles of the internal three-fold and $\hat{D}_i$ the corresponding Poincar\'e dual two-form.
We shall always project out all the $G$-moduli by considering orientifold
projections such that $h^{1,1}_- = 0 \Rightarrow h^{1,1}_+ =
h^{1,1}$.

The tree-level 4D K\"ahler potential takes the form:
\be
  K_{tree}=-2\ln \vo -\ln
  \left(S+\bar{S}\right) -\ln \left( -i\int\limits_X \Omega \wedge
  \bar{\Omega}\right)\,. \label{eqtree}
\ee
It depends implicitly on the complex structure moduli via the holomorphic (3,0)-form $\Omega$ and on the K\"ahler moduli via the Calabi-Yau volume $\vo$ expressed in units of the string length $\ell_s = 2\pi \sqrt{\alpha'}$.
The volume $\vo$ is measured by an Einstein frame
metric $g^{\scriptscriptstyle E}_{\mu \nu} = e^{-\phi/2} \,
g^s_{\mu \nu}$ and reads:
\be
  \vo = \frac 16 \int_X J\wedge J\wedge
  J = \frac 16 \, k_{ijk} t^i t^j t^k\,, \label{Vol}
\ee
where we expanded the K\"ahler form $J$ in a basis $\{ \hat{D}_i \}_{i=1}^{h^{1,1}}$ of $H^{1,1}(X,\mbb{Z})$
as $J = t^i \hat{D}_i$ and we denoted the triple intersection numbers of $X$ by $k_{ijk}$.
The volume can then be obtained as a function of the $\tau_i$ by inverting the following relations:
\be
 \tau_i  = \frac 12\int_X \hat{D}_i\wedge J\wedge J
  =\frac{\partial \vo}{\partial t^i}
  =\frac 12\, k_{ijk}\, t^j\, t^k\,. \label{TauDef}
\ee
By turning on background fluxes of the form $G_3 = F_3 +iS
H_3$, where $F_3= dC_2$ and $H_3=dB_2$, a superpotential of the following form is generated \cite{GVW}:
\be
  W_{tree}=\int\limits_X G_3 \wedge \Omega \,.
  \label{Wtree}
\ee
The fact that $W_{tree}$ does not depend on the K\"ahler moduli
combined with the no-scale structure of $K_{tree}$ implies that
only the dilaton and the complex structure moduli can be fixed at
tree-level by imposing vanishing F-term conditions \cite{gkp}.
Therefore, the $T$-moduli remain precisely massless at leading semiclassical order.

\subsection{Problems with K\"ahler moduli stabilisation}
\label{TheProblems}

In order to develop a potential for the K\"ahler moduli, one
has to consider either D-terms or corrections beyond the tree-level approximation
of the F-term potential. When one studies the effect of the corrections to the
leading approximation, the $S$ and $U$-moduli can be considered as fixed at their
flux-stabilised value. Therefore, in the study of K\"ahler moduli stabilisation we
shall consider the tree-level superpotential simply as a flux-dependent constant $W_0 = \langle W_{tree} \rangle$,
while the tree-level K\"ahler potential will take the form $K=-2\ln\vo$ with the $S$ and $U$-dependent
part just entering the F-term potential as an overall normalisation factor.

When trying to fix the K\"ahler moduli, one generically faces several problems.
Let us list the main ones:
\begin{itemize}
\item[$\triangleright$] \textit{Control over the effective field theory}:
Due to the fact that the potential for the $T$-moduli is completely flat at tree-level,
one has, in principle, to consider any possible perturbative
and non-perturbative correction to the leading order expressions, in order to lift these directions. Therefore, it is in
general difficult to have full control over the effective field theory: One has to make sure that there are no other corrections which are larger or comparable to the stabilising effects
under consideration. Two basic requirements to trust the effective field theory approach are
the weak coupling limit, i.e.\ $g_s\ll 1$, and the large volume limit, i.e.~$\vo\gg 1$.

\item[$\triangleright$] \textit{Tension between K\"ahler moduli stabilisation via non-perturbative effects and chirality}:
The authors of \cite{blumenhagen} pointed out that there is a problem with any
stabilisation technique which relies on non-perturbative
effects to fix the four-cycle supporting visible chiral matter.
The generic presence of chiral intersections between the instanton and visible sector divisors
induces a prefactor for the non-perturbative superpotential which depends on chiral matter VEVs.
In order not to break any visible sector gauge symmetry, the VEVs of these fields have
to vanish, killing the instanton contribution to the superpotential.

\item[$\triangleright$] \textit{Tension between K\"ahler moduli stabilisation via non-perturbative effects and the
cancellation of Freed-Witten anomalies}: The cancellation of Freed-Witten (FW) anomalies
requires to turn on half-integral gauge fluxes on any divisor $D$ which is \textit{non-spin},
i.e.\ with odd first Chern class $c_1(D)$ \cite{MM,FW}. The presence of non-vanishing gauge fluxes creates a
problem for all the stabilisation mechanisms which rely on more than one non-perturbative
effect to fix the K\"ahler moduli \cite{BBGW,CKMW}. In fact, in the simplest fluxless case, an
$O(1)$ instanton contributing to the superpotential is obtained by considering
a Euclidean D3-brane wrapping a rigid four-cycle which is transversally invariant under the orientifold action. The presence of possible FW fluxes would then render the instanton configuration not
invariant under the orientifold anymore. This can be cured by
compensating these half-integral fluxes by appropriately adjusting the $B$
field so that the combination $\mc{F}=F-B$ is still vanishing.
However, once the $B$-field is fixed,
it is in general not able anymore to compensate the FW fluxes of any other instantons,
killing their simultaneous contribution to the superpotential.

\item[$\triangleright$] \textit{D-term problem}: In Type IIB flux compactifications
the GUT- or MSSM-like visible sector is built via stacks of space-time filling
D7-branes wrapping divisors $D_i$ of the internal manifold. These divisors are chosen to be rigid in order to avoid unwanted matter
in the adjoint representation. Chiral matter is then obtained at the intersection with
a second stack of D7-branes via turning on an internal gauge flux.
In turn, this gauge flux generates a Fayet-Iliopoulos (FI) term $\xi_i$ which
depends on the $T$-moduli and takes the form \cite{FI}:
\be
\xi_i=\frac{1}{4\pi\vo}\int_X \hat{D}_i \wedge J\wedge \mc{F}_i
=\frac{1}{4\pi\vo}\,q_{ij}\,t^j\,,
\label{FI}
\ee
where $q_{ij}= \tilde{f}_i^k k_{ijk}$ is the $U(1)$-charge of the K\"ahler modulus $T_j$
induced by the magnetic flux $\mc{F}_i=\tilde{f}_i^k \hat{D}_k$ on $D_i$.
Including also possible matter fields $\varphi_j$ with charges $c_{ij}$ under the $U(1)$ on $D_i$,
the total D-term potential reads:
\be
V_D = \frac{g_i^2}{2} \left( \sum_j c_{ij} |\varphi_j|^2 -\xi_i\right)^2\,.
\label{VD}
\ee
This D-term potential is the leading effect to fix the volume $\tau_{\rm vis}$ of the four-cycle supporting the visible
sector. In fact, as we have already pointed out,
it is very hard to fix this modulus via non-perturbative effects
while,
as we shall see later on, any kind of perturbative correction is suppressed
with respect to $V_D$ by inverse powers of the Calabi-Yau volume.
If there are no visible sector singlets which can get a non-vanishing VEV cancelling the FI-term,
the supersymmetric locus $V_D=0$ corresponds to $\xi_i=0$, $\forall i$
\footnote{In the presence of singlets with non-zero VEV, the visible sector modulus might be fixed using
$g_s$ corrections to the K\"ahler potential \cite{LVScond}. We shall however not
consider this option in order to realise a more model-independent solution.}.
Using the classification of rigid divisors in terms of their shrinkability properties
presented in \cite{CKM}, it is then easy to show that the vanishing FI-term
requirement generically forces one or more rigid four-cycles to
shrink to zero size\footnote{See \cite{BBGW,CKMW} for explicit
examples featuring this problem.}. This causes a `D-term problem' due to the
poor control over $\alpha'$ and quantum corrections at the singular regime\footnote{Ignoring these control issues, the D-term induced shrinking of a diagonal del Pezzo
would naturally lead to quiver constructions \cite{CMQ,Quiver}.}.

\item[$\triangleright$] \textit{Stabilisation within the K\"ahler cone and phenomenological requirements}: In order to
have a positive definite metric, the K\"ahler moduli have to satisfy:
\be
\int_C J >0\,, \qquad \int_S J\wedge J>0\,, \qquad \int_X J\wedge J\wedge J>0\,,
\ee
for all complex curves $C$ and surfaces $S$ on the Calabi-Yau $X$. These conditions
define a subset of $\mbb{R}^{h^{1,1}}$ called the K\"ahler cone.
Any viable stabilisation mechanism has to fix the K\"ahler moduli
within the K\"ahler cone. However, it is in general not very straightforward
to satisfy this condition, especially if it is combined with the requirement
of obtaining phenomenologically viable scales (like GUT-scale strings or TeV-scale SUSY)
and, at the same time, the correct size of the visible sector gauge coupling. This is
given by the volume $\tau_{\rm vis}$ of the divisor $D_{\rm vis}$
supporting the GUT- or MSSM-like D7-stack plus a positive
flux dependent shift:
\be\label{alphavis}
\alpha_{\rm vis}^{-1}= \tau_{\rm vis}- \frac{1}{2 g_s}\int_{D_{\rm vis}}\mc{F}_{\rm vis}
\wedge \mc{F}_{\rm vis}\simeq \mc{O}(10 - 100)\,.
\ee
\end{itemize}

\subsection{Solutions for K\"ahler moduli stabilisation}

In this section we shall outline our general strategy to fix all the
K\"ahler moduli without facing any of the generic problems described in the previous section.
In section \ref{TheExample} we shall then illustrate our general claims in an
explicit example of a K3-fibred Calabi-Yau three-fold with del Pezzo divisors taken
from the list of \cite{CKM}.

We now explain how we intend to overcome the main challenges of K\"ahler moduli stabilisation
by going again through the list of section \ref{TheProblems}.

\subsubsection{Control over the effective field theory}

We shall assume that the background fluxes can be tuned so to obtain a value of the
dilaton that leads to the weak coupling limit $g_s= 1/{\rm Re}(S)\ll 1$.
This guarantees that perturbation theory does not break down and our Type IIB
approximation is trustworthy.

We now analyse the regime of validity of the various perturbative and
non-perturbative corrections to the leading order tree-level expressions.

\subsubsection*{Non-perturbative corrections}

Due to the non-renormalisation theorem, the first corrections to the superpotential
arise non-perturbatively and take the form \cite{kklt}:
\be
  W_{\rm np} = \sum\limits_i A_i\, e^{- a_i T_i}\,.
\label{Wnp}
\ee
The real part of the modulus $T_i$ parameterises the volume
of an internal four-cycle wrapped by either an ED3 instanton (in which case $a_i=2\pi $)
or by a stack of D7-branes supporting a condensing gauge theory
(for which $a_i=6\pi /b_0$ with $b_0$ being the coefficient of the one-loop beta function).
The threshold effects $A_i$ can be considered as $\mc{O}(1)$ constants since
they depend on the complex structure moduli which are flux-stabilised at tree-level.
We shall neglect the effect of multi-instanton contributions since we will always fix
each K\"ahler modulus such that $a_i\tau_i\gg 1$.

\subsubsection*{Perturbative corrections}

Due to the absence of a non-renormalisation theorem,
the K\"ahler potential receives perturbative corrections both in $\alpha'$ and $g_s$.
Let us analyse these two different kinds of corrections separately.

\

\noindent
{\it $\alpha'$ corrections} \\
These are higher derivative corrections which vanish in the limit $\alpha'\to 0$.
Given that in this limit also the string length $\ell_s=2\pi\sqrt{\alpha'}\to 0$,
these effects are reflecting the fact that we are dealing with one-dimensional fundamental objects
instead of ordinary point particles. The leading $\alpha'$ correction
to the 4D tree-level K\"ahler potential comes from the 10D $\alpha'^3 \mc{R}^4$ term and
behaves as \cite{bbhl}:
\be
  K= -2\ln \left( \vo+\frac{\xi }{2 g_s^{3/2}}\right)\simeq -2\ln\vo-\frac{\xi}{g_s^{3/2}\vo} \,,
 \label{KalphaP}
\ee
where the parameter $\xi =-\frac{\chi (X)\zeta (3)}{2(2\pi )^3}$
depends on the Calabi-Yau Euler number $\chi(X)$
and the Riemann zeta function $\zeta(3) \simeq 1.2$.
The $\alpha'$ corrections are given by an expansion
in inverse powers of the overall volume,
and so we can focus just on the leading order effect (\ref{KalphaP})
only if $\vo \gg \xi/g_s^{3/2} \gg 1$. This condition will
definitely be satisfied since we shall fix the
Calabi-Yau volume exponentially large in string units.

\

\noindent
{\it $g_s$ corrections} \\
The one-loop open string corrections to the K\"ahler potential generically take the form \cite{bhk}:
\be
 \delta K_{(g_s)}=\delta K_{(g_s)}^{KK}+\delta K_{(g_s)}^W\,,
\ee
where, in the closed string channel, $\delta K_{(g_s)}^{KK}$ can
be interpreted as coming from the exchange between non-intersecting D7-branes
of Kaluza-Klein strings\footnote{In general, these corrections come also from
the exchange of Kaluza-Klein strings between D7 and D3-branes (and O3-planes) but,
as we shall see later on, in our case
tadpole cancellation does not force us to introduce any D3-brane. This is indeed good news
since we do not have additional moduli and we do not lower the D3-charge of background fluxes.},
while $\delta K_{(g_s)}^W$ is due to exchange of winding strings between
intersecting stacks of D7-branes (or intersecting D7-branes and O7-planes).

The authors of \cite{bhp} proposed a generalisation to the case of Calabi-Yau three-folds
of the explicit computation of these $g_s$ effects for simple toroidal orientifolds \cite{bhk}.
It has been possible to conjecture the behaviour of these string loop corrections
because of their simple dependence on the K\"ahler moduli and
their interpretation as the tree-level propagation of closed strings.
Moreover, this conjecture has passed several low energy tests since
it reproduces the correct behaviour of the Coleman-Weinberg potential \cite{ccq}.
The result for the behaviour of these loop corrections is:
\be
 \delta K_{(g_s)}^{KK}\sim \sum\limits_i
 \frac{c_i^{KK}(U,\bar{U})\,m_{KK}^{-2} }{\hbox{Re}
 \left( S\right) \vo} \sim \sum\limits_i
 \frac{c_i^{KK}(U,\bar{U})\, t^{\perp}_i
 }{\hbox{Re} \left( S\right) \vo},  \label{KgsKK}
\ee
where $t^{\perp}_i$ is the volume of the two-cycle transverse to the two non-intersecting D7-branes, and:
\be
 \delta K_{(g_s)}^W\sim \sum\limits_i
 \frac{c_i^W(U,\bar{U})\, m_W^{-2}}{\vo}\sim
 \sum\limits_i\frac{c_i^W(U,\bar{U})}{\vo\,t^{\cap}_i}, \label{KgsW}
\ee
with $t^{\cap}_i$ now being the volume of the two-cycle where the two D7-branes
intersect (or the D7-brane and the O7-plane intersect).
In what follows, we shall always check that each two-cycle
volume is fixed larger than unity in order to avoid the
blow-up of these string loop corrections.
Notice that $c^{KK}_i $ and $c^W_i$ are unknown
functions of the complex structure moduli, which may be simply
regarded as unknown constants for the present purposes because the
$U$-moduli are already flux-stabilised by the
leading-order dynamics. The KK correction $\delta K_{(g_s)}^{KK}$ (\ref{KgsKK})
dominates over the $\alpha'$ correction (\ref{KalphaP}) in the large volume
limit where we can trust the effective field theory. However, as noticed in \cite{bhp,ccq},
the leading order contribution of $\delta K_{(g_s)}^{KK}$
to the scalar potential is vanishing, producing a subleading effect that is
subdominant relative
to the leading $\alpha'$ correction in the large volume limit.
This `extended no-scale structure' will allow us to study moduli stabilisation
in two steps: first working at leading order where we shall neglect the
$g_s$ corrections and focus only on the interplay between
non-perturbative and $\alpha'$ effects, and then studying the potential at subleading
order where we shall show how the string loop corrections can lift any remaining flat direction.

\subsubsection{K\"ahler moduli stabilisation and chirality}

Following \cite{CKM,BBGW}, there are two interesting different classes of rigid divisors:

\begin{itemize}
\item \emph{`Diagonal' del Pezzo}: there exists a basis of toric divisors comprising this del Pezzo four-cycle which appears in the intersection form in a diagonal way.
On a Calabi-Yau three-fold, del Pezzo submanifolds are arbitrarily contractible to a point without affecting
the rest of the geometry. For `diagonal' del Pezzo divisors this becomes obvious, since their volumes are as in (\ref{TauDef}) but with $k_{ijk}\neq 0$ only if $i=j=k$.

\item \emph{Rigid but not del Pezzo}: for any choice of basis of four-cycles, these rigid divisors appear in the volume form in a non-diagonal way. Moreover, they are not contractible to a point.
\end{itemize}

As pointed out in \cite{CKM}, the natural candidate for supporting non-perturbative effects
is a `diagonal' del Pezzo divisor. In the presence of intersecting
branes, we shall perform a choice of fluxes on them
which guarantees the absence of any chiral intersection between the hidden sector,
responsible for the generation of the non-perturbative effects,
and the visible sector D7-stack. In this way, we shall solve
the tension between K\"ahler moduli stabilisation and chirality\footnote{Another solution
relies on the possibility to consider compactifications with $h^{1,1}_-\neq 0$ \cite{Grimm}.}.

Notice that we shall also carefully choose a flux configuration
that gives rise to a vanishing total gauge flux on the hidden sector
D7-stack. This is important to ensure that the hidden sector is a pure
supersymmetric Yang-Mills theory (that definitely undergoes gaugino
condensation) and, above all, that no FI-term gets generated.
Otherwise the presence of a non-zero flux would force this `diagonal' del Pezzo four-cycle to shrink to zero size.

The reason why we shall focus on gaugino condensation instead of ED3 instantons
is purely phenomenological. In fact, we shall fix the overall volume by
the interplay of non-perturbative and $\alpha'$ corrections at:
\be
\vo\sim W_0\,e^{ a \tau_{\rm diag}/g_s}\,, \label{IlV}
\ee
where $\tau_{\rm diag}$ denotes the volume of the `diagonal' del Pezzo
in string frame while, as we have seen before, in the case of an ED3 instanton,
$a=2\pi$. Therefore, the two conditions $\tau_{\rm diag}\gtrsim 1$ and $g_s\lesssim 0.1$,
needed to trust the effective field theory, imply $\vo/W_0 \gtrsim 10^{27}$.
In turn, this sets a very constraining upper bound on the gravitino mass:
\be
m_{3/2}= e^{K/2} W_0 M_P = \sqrt{\frac{g_s}{8\pi}} \frac{W_0 M_P}{\vo} \lesssim 55 \,{\rm meV}\,,
\ee
which is clearly incompatible with gravity mediated supersymmetry breaking.
Moreover, for natural values of $W_0\sim\mc{O}(1)$, the string scale would
also turn out to be very low:
\be
M_s= \frac{M_p}{\sqrt{4\pi\vo}}\lesssim 15 \,\textrm{TeV},
\ee
forbidding the possibility of standard GUT theories. On the other hand,
in the case of gaugino condensation, $a=6\pi/b_0$, and so we have an
additional parameter, $b_0$, which can be varied. This gives us more freedom to obtain the desired phenomenological
scales (like GUT-scale strings and TeV-scale SUSY)\footnote{See also \cite{Reheating} for a similar line of argument.}. However
we stress that $b_0$ cannot be varied arbitrarily since its value
will be constrained by the D7 tadpole cancellation condition.

\subsubsection*{Gaugino condensation}

Let us now discuss the form of the non-perturbative superpotential
generated by gaugino condensation. The one-loop running of the gauge kinetic function is
given by:
\be
\frac{1}{g^2}(\mu)=\frac{1}{g_0^2}-\frac{b_0}{16 \pi^2}\ln\left(\frac{\Lambda_{\rm UV}^2}{\mu^2}\right),
\label{couplN}
\ee
where $\mu$ is a generic energy scale, $\Lambda_{\rm UV}$ the UV scale
and $g_0$ the bare coupling.
In our constructions, the `diagonal' del Pezzo supporting the non-perturbative effects
is transversally invariant under the orientifold projection.
Therefore, the field theory living on this four-cycle is a pure $Sp(2 N_{gc})$ theory that is known to undergo gaugino condensation at the
scale $\mu=\Lambda_{\rm IR}$ where the expression (\ref{couplN}) formally diverges, i.e.\ at:
\be
\Lambda_{\rm IR}=\Lambda_{\rm UV}\,e^{-\frac{8 \pi^2}{b_0\,g_0^2}}.
\ee
The non-perturbative superpotential which gets generated looks like:
\be
W_{\rm np} = \Lambda_{\rm IR}^3 = \Lambda_{\rm UV}^3\,e^{-\frac{3}{b_0}\frac{8 \pi^2}{g_0^2}}.
\label{strongWN}
\ee
In string theory we have that $g_0^2=4\pi/\tau_{\rm diag}$ without any dilaton-dependent correction since there is no flux turned on, and
the UV scale of the 4D supergravity is the Planck scale $\Lambda_{\rm UV}= M_P$.
Therefore introducing a generic prefactor $A$, which we expect to be of $\mc{O}(1)$, we obtain
(setting $M_P=1$ and complexifying the K\"ahler modulus):
\be
W_{\rm np} = A \,e^{-\frac{3}{b_0}\,2\pi \,T_{\rm diag}}.
\label{strongWnpN}
\ee
In the case of a pure $SU(N_{gc})$ theory $b_0= 3\,N_{gc}$, so we would obtain $W_{\rm np} = A \,e^{-\frac{2\pi T_{\rm diag}}{N_{gc}}}$,
while in the case of a pure $Sp(2N_{gc})$ theory $b_0= 3\left(N_{gc}+1\right)$ \cite{Intriligator}, leading to:
\be
W_{\rm np} = A \,e^{-\frac{2\pi T_{\rm diag}}{N_{gc}+1}}.
\ee

\subsubsection{K\"ahler moduli stabilisation and Freed-Witten anomalies}

We shall solve the tension between fixing
more than one K\"ahler modulus by non-perturbative effects and
the cancellation of FW anomalies, via exploiting a
moduli stabilisation technique which relies on the
the existence of just one non-perturbative effect.
This is the so-called LARGE Volume Scenario \cite{LVS} where the overall
volume is fixed by the interplay of the leading order $\alpha'$
correction and the non-perturbative effects supported by
a single `diagonal' del~Pezzo divisor\footnote{
Another solution relies on the possibility to consider compactifications with a rigid ample divisor \cite{Bobkov}.} \cite{LVScond}.
In the next section,
we shall explain how to fix most of the remaining moduli
at leading order via D-terms without hitting the walls
of the K\"ahler cone. If some flat directions are still left over,
they will be lifted by subleading string loop corrections.

As we have already said, we shall consider a confining hidden sector
given by a stack of D7-branes wrapping a transversally invariant rigid four-cycle without
chiral intersections with the visible sector. Thus, we obtain an $Sp(2N_{gc})$ group
with the presence of an FW half-integer. This flux would break the symplectic group to an
unitary one and generate dangerous chiral matter that might kill
gaugino condensation. Hence, we shall cancel this FW flux by an appropriate
choice of the $B$ field.

\subsubsection{D-term problem}

Let us now explain a generic strategy to avoid the D-term induced shrinking of the visible sector four-cycle.

If the visible sector is supported by a `diagonal' del Pezzo then the FI-term simplifies considerably to $\xi = c \sqrt{\tau_{\rm dP}}$, where $c$ is a
flux-dependent non-zero constant. Hence, the requirement $\xi=0$ forces $\tau_{\rm dP}$
to shrink to zero~\cite{BBGW,CKMW,CKM}.
On the other hand, if the visible sector D7-stack wraps
a `non-diagonal' rigid divisor
(like in the last two cases of the previous classification)
which intersects other $(n-1)$ four-cycles, the corresponding
FI-term depends linearly on $n$ K\"ahler moduli. Imposing the vanishing of this FI-term fixes the corresponding combination of K\"ahler moduli.
This observation
gives some hope to fix the moduli without going to the singular regime.
However, if also all the other $(n-1)$ divisors support a $U(1)$ gauge
theory, then the simultaneous vanishing of all the FI-terms
gives rise to a set of $n$ linear homogeneous equations in $n$ unknowns which
admits only the vanishing solution if the rank $r$ of the corresponding
flux-dependent matrix is $n$. Hence, one could still face the
shrinking problem even if the visible
sector D7-stack is placed on a `non-diagonal' rigid divisor.
Therefore, in order to avoid the shrinking solution, we have to make sure that the rank $r$ of the D-term conditions is smaller than the number $n$ of entering K\"ahler moduli (i.e.\ $r<n$).
This implies that the D-terms cannot fix all the moduli. They will leave $(n-r)$ flat directions.

There are two ways to get $r<n$:
\begin{enumerate}
\item By construction, i.e.\ by wrapping $p<n$ divisors;
\item By an appropriate choice of fluxes which renders some FI terms parallel.
\end{enumerate}

In our constructions the modulus controlling the size of the `diagonal' del~Pezzo does not enter in the D-term constraints (and will be fixed by non-perturbative effects). On the other hand, in general the FI-terms will depend on the other $n=h^{1,1}-1$ moduli.

The condition of reproducing the correct size of the visible sector gauge coupling
$\alpha^{-1}_{\rm vis}$ forces the divisor wrapped by the GUT- or MSSM-like D7-stack
to be fixed relatively small since too large values of $\tau_{\rm vis}$
would give rise to too small values of $\alpha^{-1}_{\rm vis}$ (see \eqref{alphavis}).
This cannot be reached if we have $n-r=1$ and on top of this we require to realise the LVS. In fact, in order to have LVS we need that at least one K\"ahler modulus is exponentially large. On the other hand $n-r=1$ implies that all the K\"ahler moduli are of the same size (as we have $n-1$ linear equations in $n$ unknowns). Therefore,
if we want to obtain the LVS (at least one large modulus) and realistic gauge coupling (one small modulus), we need $n-r>1$. Hence, if we set $r=1$, we have to work with $n=3$.
In terms of the total number of K\"ahler moduli,
we are left over with the minimal set-up with $h^{1,1}=n+1=4$.
In this case we would also obtain enough intersecting cycles to construct a GUT- or MSSM-like
chiral model. We finally stress again that if $n=r$ we would be driven to the quiver locus by the D-term conditions.

It remains to study how to stabilise the unfixed $(n-r)$ K\"ahler moduli.
The minimal number of flat directions left over by the D-terms is $n-r=1$
for cases where we allow the four-cycles to have the same size (and so no LVS is possible) while $n-r=2$ for
models where one direction can be sent large (like in the LVS). We point out that,
due to the tension between non-perturbative effects and FW anomaly cancellation,
we would like to fix these $n-r$ directions without using non-perturbative effects since
gaugino condensation has already been used to fix the `diagonal' del Pezzo. Hence, we need
to look at perturbative corrections to the K\"ahler potential: one flat direction
can be fixed by $\alpha'$ corrections whereas the other, if present, can be
fixed by subleading string loop effects.

\subsubsection{Stabilisation within the K\"ahler cone}

In the next section we will present different chiral models
which illustrate our general claims in the case of a K3-fibred
Calabi-Yau three-fold with a `diagonal' del Pezzo divisor and $h^{1,1}=4$.
This example features also an interesting F-theory uplift
in terms of an elliptically fibred Calabi-Yau four-fold and
it is taken from the list of \cite{CKM} consisting
of 158 examples of lattice polytopes
which have a Calabi-Yau hypersurface with $h^{1,1}=4$
that admits a K3 fibration structure and at least one `diagonal' del Pezzo.

We will make three explicit choices of brane set-up and fluxes
where the underlying parameters ($g_s$, $W_0$ and $A$) can be chosen consistently with K\"ahler moduli stabilisation
within the K\"ahler cone and
the regime of validity of the 4D effective field theory. At the same
time, the same choice yields also interesting phenomenological scales
and the correct order of magnitude of the visible sector gauge coupling.

In all cases we set the number of intersecting K\"ahler moduli
$n=3$.
In the first case we shall generate just $r=2$ D-term conditions
by wrapping two divisors (in addition to the `diagonal' del~Pezzo) on which we will switch on fluxes. We shall
then lift the remaining flat direction with $\alpha'$ corrections
obtaining a Calabi-Yau volume of the order $\vo\sim 5\cdot 10^3$
which gives rise to GUT-scale strings and TeV-scale SUSY by fine-tuning
$W_0\sim 5\cdot 10^{-9}$.

In the last two cases we shall generate just $r=1$ D-term condition
by wrapping the same number of divisors as above, but making
a proper choice of fluxes which renders the two remaining FI-terms parallel.
We shall then lift the two remaining flat directions with both $\alpha'$ and $g_s$ corrections
obtaining an exponentially large internal volume.
In the second case, corresponding to $\vo \sim 10^{12}$,
we obtain TeV-scale SUSY without fine-tuning the background fluxes, i.e.\ $W_0\sim\mc{O}(1)$,
while in the third case, corresponding to $\vo\sim 10^{29}$,
we get TeV-scale strings for a very anisotropic configuration with four
small and two micron-sized extra dimensions.

\section{A global chiral model for a K3-fibred Calabi-Yau}
\label{TheExample}

In this section we shall construct a global model with chiral matter.
We will start by giving the details of the compactification manifold we have chosen.
Then, we will describe all the consistency constraints that the model has to satisfy.

\subsection{Geometric data of the Calabi-Yau three-fold}

In \cite{CKM} a search for Calabi-Yau three-folds that are K3 fibrations was performed.
In order to construct our model, we have picked one among them.

The Calabi-Yau three-fold $X$ is a hypersurface in a four-dimensional toric ambient variety. This toric four-fold is given by its simplicial fan which is constructed out of a four-dimensional reflexive lattice polytope and one of its sixteen maximal coherent surface-triangulations. These data are also encoded in the following weight matrix and Stanley-Reisner ideal:
\be\label{TableWeights}
 \begin{array}{|c|c|c|c|c|c|c|c||c|}
 \hline  z_1 & z_2 & z_3 & z_4 & z_5 & z_6 & z_7 & z_8 & D_\textmd{X} \tabularnewline \hline \hline
    1  &  1 &   1 &   0 &   0 &   0 &   1 &   4  & 8 \tabularnewline\hline
    1  &  1 &   0 &   0 &   0 &   1 &   0 &   3  & 6 \tabularnewline\hline
    0  &  1 &   1 &   1 &   0 &   0 &   0 &   3  & 6 \tabularnewline\hline
    0  &  1 &   0 &   0 &   1 &   0 &   0 &   2  & 4 \tabularnewline\hline
 \end{array}\,,
\ee
\be\label{eq:SR-ideal}
\mbox{SR} = \{z_2 z_5, z_1 z_6, z_1 z_7, z_5 z_7, z_2 z_4 z_6,  z_3 z_4 z_8, z_3 z_7 z_8  \}\,.
\ee
From the weight matrix we obtain the four equivalence relations for the homogeneous coordinates. The SR-ideal tells us which homogeneous coordinates are not allowed to vanish simultaneously. The column $D_X$ in~\eqref{TableWeights} indicates the multi-degrees of the Calabi-Yau hypersurface divisor $X$ such that we obtain a manifold with a vanishing first Chern class. The Hodge numbers of the Calabi-Yau $X$ are $h^{1,1}=4$ and $h^{1,2}=106$, and consequently Euler characteristic $\chi$ is equal to $2\left(h^{1,1}-h^{1,2}\right)=-204$.
All this information has been obtained by means of PALP~\cite{PALP}.

Having the full toric data of the ambient space, we can reduce its intersection ring to the Calabi-Yau three-fold. Using the algorithm presented in~\cite{diss,CKM}, we are led to choose the following basis for $H^{1,1}(X,\Zbb)$:\footnote{Since the intersection polynomial is only semi-diagonalisable, the choice of basis is not canonical. However, this one is special in the sense that it will allow to simplify many expressions.}
\be
 \G_1 = D_7\,,  \qquad \G_2 = D_2 + D_7\,,  \qquad \G_3 = D_1\,, \qquad \G_4 = D_5 \,.
\label{newBasN}
\ee
The intersection polynomial takes the form:
\be\label{IntersForm}
 I_3 = 2 \G_1^3 +4\G_2^3 +4\G_4^3 +2 \G_2^2\G_3 -2 \G_4^2\G_3\,.
\ee
The $\G$'s are an integral basis, i.e.~all integral cycles can be written as integral linear combination of these basis elements.
In terms of the chosen basis, the divisors $D_i$ are given by:
\begin{align}
 D_1 &=  \G_3, & D_2 &= \G_2- \G_1,  & D_3&= \G_2-\G_3 -\G_4,  \nonumber\\
 D_4 &= \G_2-\G_1 -\G_3 -\G_4,  & D_5& = \G_4, & D_6&=\G_3-\G_1, \\
 D_7 &=\G_1, & D_8& = 3\G_2-2\G_1 -\G_4, \nonumber
\end{align}
For completeness we report also the second Chern class of the Calabi-Yau:
\be
 c_2(X) = 4\G_1^2 + 12\G_2^2 -2 \G_3^2  -16 \G_1\G_2  +5 \G_1\G_4 +2 \G_2\G_3 - 7 \G_2\G_4 -2 \G_3\G_4   \:.
\ee
As we have said, the Calabi-Yau $X$ is a K3 fibration with\footnote{The lattice polytope of the ambient space given in~\eqref{TableWeights} has the $\mathbb Z_2$ symmetry $z_1\leftrightarrow z_3,\,z_4\leftrightarrow z_6$.  This remains true also on the level of the Calabi-Yau, since, if we reduce~\eqref{eq:SR-ideal} to the hypersurface also the SR-ideal respects this symmetry. 
This implies that there is a second equivalent K3 fibration given by the projection \eqref{eq:k3-projection} with $z_1\leftrightarrow z_3,\,z_4\leftrightarrow z_6$.}
\be\label{eq:k3-projection}
\pi:\quad [z_1\,:\,z_2\,:\,z_3\,:\,z_4\,:\,z_5\,:\,z_6\,:\,z_7\,:\,z_8]\mapsto [z_1\,:\,z_6 z_7]\,.
\ee
One of the  K3 fibres is given by the divisor $D_1$. Moreover, $X$ has one dP$_7$, corresponding to the divisor $D_7$. As one can sees from \eqref{IntersForm}, this is a `diagonal' del~Pezzo.
Its Euler characteristic is $\chi (D_7)=10$ and $h^{1,1}(D_7)=8$.
There are three other rigid, but non del Pezzo, four-cycles: $D_5$ with Euler characteristic $\chi (D_5) = 8$ and $h^{1,1}(D_5)=6$ \footnote{Analysing the toric data, one finds that the four-cycle $z_5=0$ is actually a ruled surface.}, and
$D_4$ and $D_6$ with $\chi (D_4)=\chi (D_6)=14$ and $h^{1,1}(D_4)=h^{1,1}(D_6)=12$. All of them have $h^{0,1}=h^{0,2}=0$.

Expanding the K\"ahler form in the basis $\{\Gamma_i\}_{i=1}^4$ as $J= \sum_i t_i \,\G_i$,
we can express the volume of $X$ in terms of the K\"ahler moduli $t_i$:
\be
 \vo = \frac13 \left[ 2 t_2^3 +3 t_2^2 t_3  + t_4^2( 2 t_4-3 t_3)+t_1^3\right]\,.
\label{VolCY}
\ee
The volumes of the relevant divisors are given by:
\bea
 \tau_1 &=& (t_2 -t_4) (t_2 + t_4) \:, \nonumber \\
 \tau_4 &=& (t_2 + t_4)(t_2+2 t_3-t_4)-t_1^2 \:,\nonumber\\
 \tau_5 &=&  2 t_4 (t_4 - t_3) \:, \\
 \tau_7 &=&  t_1^2 \:.\nonumber
\eea
In order to know the parameter range of the coefficients $t_i$, we need the K\"ahler cone. This is the dual to the Mori cone, i.e.\ the cone of effective curves. The effective curves of the Calabi-Yau hypersurface are usually not straightforwardly obtained. So one could try to approximate the Mori cone of the hypersurface by the Mori cone of the ambient space. However, we have to take into account that there may be flop transitions in the ambient space which do not affect the Calabi-Yau~\cite{Batyrev,Wall1966}. Hence, to compute the K\"ahler cone of the Calabi-Yau one has to take into account the cones of these different ambient spaces giving the same Calabi-Yau three-fold.
For the example at hand, we find seven triangulations of the lattice polytope that lead to the same Calabi-Yau hypersurface. These triangulations are connected via flop transitions that change the triangulation of a certain facet of the lattice polytope. Under these transitions some of the effective curves of the toric ambient space become non-effective. However, these curves do not intersect the hypersurface and there is no phase transition on the Calabi-Yau. Therefore, the effective curves relevant for the hypersurface are those that are effective in all these triangulations. The dual cone to this intersection of Mori cones is the following:
\be
r_1\equiv - t_1>0\,, \qquad r_2\equiv t_1 + t_2 + t_4>0\,, \qquad r_3\equiv t_3 - t_4>0\,, \qquad r_4\equiv-t_4>0\,,
\label{Kcone}
\ee
which is the (approximate) K\"ahler cone of the Calabi-Yau hypersurface\footnote{An explicit computation shows that this cone is actually the union of the cones descending from the seven different triangulations of the ambient space.}.

As explained above, we shall place the hidden sector supporting gaugino condensation
on the `diagonal' del Pezzo divisor $D_7$ whose volume, $\tau_7=t_1^2$, appears in
the overall volume (\ref{VolCY}) in a diagonal way. We can also see explicitly from
the K\"ahler cone conditions (\ref{Kcone}) that $t_1$ can be shrunk to zero without affecting the other cycles.

\subsection{Orientifold projection and D7-brane configuration}

We want to construct globally consistent models on D7-branes. Since we want the background to
preserve ${\cal N}=1$ supersymmetry in four dimensions and we want to cancel the D7-charge of the D7-branes,
we consider an orientifold projection that allows for O7 and O3 planes. It takes the form ${\cal O}=(-1)^F \Omega_p \,\sigma$, where
$\sigma$ is a holomorphic involution of the Calabi-Yau three-fold. The involution $\sigma$ acts on the K\"ahler form as $J\rightarrow J$ and on the
holomorphic three-form as $\Omega\rightarrow -\Omega$.

We choose the following involution:
\be\label{ORpr149}
 \sigma: \qquad z_8 \mapsto - z_8 \:,
\ee
that produces an O7-plane on the divisor $D_8$.
The reason for this choice is that the divisor $D_8$ has the largest weights (see table \eqref{TableWeights}). In this way it is easier to
satisfy the D7-tadpole cancellation condition with the several D7-brane stacks one would like to use to build a chiral model.

There could also be O3-planes on points of the Calabi-Yau three-fold left unchanged by the involution $\sigma$. These points are counted by
the following intersection numbers:
\be
 D_1\cdot D_2 \cdot D_6|_X,  \,\,\, D_3\cdot D_6\cdot D_7|_X,  \,\,\,  D_1\cdot D_5\cdot D_6|_X,
 \,\,\, D_2\cdot D_3\cdot D_4|_X, \,\,\, D_1\cdot D_4\cdot D_7|_X,  \,\,\, D_3\cdot D_4\cdot D_5|_X \:.
 \nonumber
\ee
Given that all these intersections are empty, there are no O3-planes.

The chosen involution is such that $h_-^{1,1}(X)=0$, and so $h_+^{1,1}(X)=h^{1,1}(X)=4$.
The symmetric equation defining the Calabi-Yau is:
\be\label{CY3eq}
 z_8^2 = P_{8,6,6,4}(z_1,...,z_7) \,,
\ee
where $ P_{8,6,6,4}(z_1,...,z_7)$ is a polynomial of degrees $(8,6,6,4)$ in the coordinates $(z_1,...,z_7)$.

The equation \eqref{CY3eq} is in the canonical form for an F-theory uplift\footnote{
Singularities of the Calabi-Yau three-fold arise when the polynomial $P$ factorises.
Sometimes the degree of the polynomial $P$ forces a factorisation. This does not happen for the example at hand. Further, singularities can also arise when
the hypersurface hits some singularities of the ambient space. Since we consider Calabi-Yaus from refelxive polytopes, this does not happen for the generic equation.
Even though \eqref{CY3eq} is symmetric under the orientifold involution and therefore not generic, nevertheless it omits all singularities.}.
The corresponding four-fold will
be an elliptic fibration over the three dimensional manifold spanned by the coordinates $(z_1,...,z_7)$.

\subsection{D7-brane stacks}

On the Calabi-Yau $X$, the total charge of a D7 brane with gauge flux $F$ is\footnote{$[D7]$ is the homology class of the four-cycle wrapped by the D7-brane. In this paper we will use the same symbol for the classes of four-cycles and their Poincar\'e dual two-forms.}
\be\label{GammaD7}
 \Gamma_{D7}
      =  [D7] + [D7]\wedge {\cal F} + [D7]\wedge\left( \frac12 \,{\cal F} \wedge {\cal F}  + \frac{c_2(D7)}{24} \right)\,,
\ee
while the charge of the O7-plane is
\be\label{GammaO7}
 \Gamma_{O7}
      =  -8[O7]  + [O7]\wedge \frac{c_2(O7)}{6} \,.
\ee
The D7-charge is given by the two-form.
In our case the D7-charge of the O7-plane is $-8D_8$. Hence, in order to satisfy the D7-brane tadpole cancellation condition,
we have to introduce D7-branes whose classes sum up to $8D_8$. If we decide to introduce only
one D7-brane, it must be described by an equation of the following form:
\be\label{WhitEq}
 \eta^2 - z_8^2 \, \chi\,,
\ee
with $\eta$ and $\chi$ generic polynomials. The reason for this form is that the D7-brane
must have double intersections with the O7-plane \cite{Collinucci:2008pf,Braun:2008ua}.
The D7-brane given by the equation \eqref{WhitEq} has singularities, around which the surface takes the
shape of the so-called Whitney umbrella. For this reason, such brane is referred to as a {\it Whitney brane}.

When $\chi$ has the non-generic form $\chi\equiv \psi^2$, the Whitney brane splits into a brane and its
image: $(\eta - z_8 \psi)(\eta + z_8 \psi)=0$. We can then understand the Whitney brane as the result of the opposite transition, i.e.
as the recombination of one D7-brane with its image.

We will not split the Whitney brane. We will consider instead factorisations of the
polynomials $\eta$ and $\chi$ like:
\be
 \eta = z_i^m \tilde{\eta}\,, \qquad\qquad \chi = z_i^{2m} \tilde{\chi}\,,
\ee
such that:
\be
 \eta^2 - z_8^2 \chi = z_i^{2m}(\tilde{\eta}^2 - z_8^2 \tilde{\chi}) \,.
\ee
The brane factorises into a Whitney brane of lower degrees along $\tilde{\eta}^2 - z_8^2 \tilde{\chi}=0$ and an $Sp(2m)$ stack along $z_i=0$. This stack is
made up of $m$ D7-branes, plus their images, along the invariant divisor $z_i=0$ (recall that $z_i\mapsto z_i$ under $\sigma$, if $i\not=8$).
Given that $D_i$ is transverse to the O7-plane, the gauge group is $Sp(2m)$ \footnote{We are using the convention in which $Sp(2)\cong SU(2)$.}.

Let us focus on our specific Calabi-Yau three-fold and analyse which brane-stacks we can factor out of the Whitney brane.
If we took just one Whitney brane to cancel the D7-tadpole introduced by the O7-plane, it would wrap the locus given by equation:
\be\label{Whit106-1}
 \eta_{16,12,12,8}^2 - z_8^2 \,\chi_{24,18,18,12} = 0\,,
\ee
where $\eta_{16,12,12,8}$ and $\chi_{24,18,18,12}$ are (non-factorised) polynomials of variables $(z_1,...,z_7)$ and of the given degrees.

In this paper we want to study two different brane configurations.
In the first case we shall wrap the rigid divisors $D_4$ and $D_5$, while in the second case,
the K3 divisor $D_1$ and the rigid divisor $D_4$ . Moreover, we will need a number of D7-branes wrapping the `diagonal' del Pezzo divisor.

Here we will analyse the generic situation in which
there are $N_{k3}$ branes (plus their images) wrapping the K3 divisor $D_1$,
$N_a$ branes (plus their images) wrapping the rigid divisor $D_4$, $N_b$ branes (plus their images)
wrapping the rigid divisor $D_5$ and $N_{gc}$ branes (plus their images) wrapping the diagonal dP$_7$ divisor $D_7$.
The results are valid for both configurations under consideration.

The chosen brane set-up requires that the equation \eqref{Whit106-1} factorises as
\be
 z_1^{2N_{k3}}z_4^{2N_{a}}z_5^{2N_{b}}z_7^{2N_{gc}} W = 0\,,
\ee
where the remaining D7-tadpole is fulfilled by a Whitney brane given by the equation:
\be\label{NewWithW}
 W \equiv \eta_{(16-N_{gc}-N_{k3},12-N_{k3},12-N_{a},8-N_{b})}^2 - z_8^2\, \chi_{(24-2N_{gc}-2N_{k3},18-2N_{k3},18-2N_a,12-2N_b)} = 0\,.
\ee
We see that, in order to keep having holomorphic (supersymmetric) D7-branes, we have to impose the following constraints on the number of branes:
\be
 N_a \leq 9\,, \qquad N_b \leq 6\,, \qquad N_{k3}\leq 9\,, \qquad N_{gc}\leq 12 - N_{k3} \,.
\ee
Since we do not want other branes apart for the ones listed, we require that the polynomial $W$ does not factorise further. This is a non-trivial requirement
in the setup we are considering. It can easily happen that asking for some values of $N_i$, the degrees in \eqref{NewWithW} force the polynomial $W$ to factorise. Let us consider a simple example to make this clear. Take the polynomial $P_{1,2,0,0}(z_1,...,z_7)$. Looking at the table \eqref{TableWeights}
we immediately see that $P$ can depend only on $z_1$, $z_6$ and $z_7$. Moreover, since the first degree is $1$, we cannot have more than one factor of $z_1$ in any monomials of $P$. This forces any monomial of $P$ to have a factor of $z_6$, leading to the factorised form $P=z_6\tilde{P}$.

A sufficient condition for this factorisation not to happen is to prove that a generic polynomial with the degrees of $W$ contains monomials that do not have a common factor. In our case, it is sufficient to prove this for $\eta$ instead of $W$ (because if $\eta$ does not factorise, $W$ does not too).
Three possible monomials with the same degrees of $\eta$ and that do not have a common factor are:
\begin{align}
 &z_1^{12-N_{k3}} z_4^{12-N_a} z_5^{8-N_b} z_7^{4-N_{gc}}\,, \\
 &z_3^{12-N_{a}} z_5^{8-N_b} z_6^{12-N_{k3}} z_7^{4-N_{gc}+N_a-N_{k3}}\,, \\
 &z_1^{4+N_a-N_{gc}-N_{k3}} z_2^{8-N_b} z_3^{4-N_{a+N_b}} z_6^{N_b+N_{gc}-N_a} \,.
\end{align}
For these to be well defined, one has the conditions
\be\label{CondNiFact}
 N_{gc}\leq 4\,, \qquad N_{gc}+N_{k3} \leq 4 + N_a\,, \qquad N_a-N_b \leq N_{gc} \,.
\ee
This is a sufficient condition for the Whitney brane $W=0$ to be non-factorised.
In particular, we note that in order to have an $SU(5)$ stack on $D_4$ (as we will do in our first explicit construction), one needs $N_b\geq 1$.

\subsection{D7 gauge fluxes and Freed-Witten anomaly}

In this section we study the gauge flux configuration paying particular attention
to the freedom left over in this choice. In fact, we are forced to
switch on particular fluxes on the various D7 stacks due to some consistency conditions.
In order to cancel the Freed-Witten anomaly, the gauge flux $F$ on the brane wrapping the divisor $D$ must satisfy \cite{MM,FW}:
\be
 F + \frac{c_1(D)}{2} \,\,\,\in\,\,\, H^2(X,\Zbb)\,.
\ee
A sufficient condition to satisfy this relation is to switch on a
flux of the form:
\be
F= \sum_{i=1}^4 f_i \Gamma_i - \frac 12\,c_1(D)  \qquad \mbox{with} \qquad f_i\in\mbb{Z}\,.
\ee
Since $D$ is embedded in a Calabi-Yau,
$c_1(D)=-D$. We will consider only fluxes that are non-trivial\footnote{Here `trivial' refers to the two-cycle that is Poincar\'e dual in $D$ to the two-form defining the flux. This two-cycle can be non-trivial in $D$ but trivial in the Calabi-Yau three-fold.} in the Calabi-Yau $X$, as the chiral intersections
depend only on them. Possible trivial fluxes would in any case contribute positively to the D3 charge, that we want to avoid, unless we need them to fix the D7-brane moduli \cite{Luca,Lust:2005bd,Braun:2008pz}.

By construction, the Whitney brane has a flux that is trivial on the Calabi-Yau $X$.
We shall consider this flux when computing the D3 charge of a Whitney brane.
Recalling that $c_1(K3)=0$, the fluxes on each D7-brane of the different stacks are then taken as follows:
\bea\label{FluxConfig}
 F_a^\sigma &=& \sum_{i=1}^4 a_i^\sigma\G_i +\frac 12 \,[D7_a]\,,\qquad
 F_b^\rho = \sum_{i=1}^4 b_i^\rho\G_i +\frac 12 \,[D7_b]\,,  \\
 F_{k3}^\mu &=& \sum_{i=1}^4 m_i^\mu\G_i \,,  \qquad
 F_{gc}^\gamma = \sum_{i=1}^4 g_i^\gamma\G_i +\frac 12\, [D7_{gc}]\,, \nonumber
\eea
where the indices $\sigma$, $\rho$, $\mu$ and $\gamma$ run over the different branes of the stack (for an $Sp(2m)$-stack, the index runs from $1$ to $m$)\footnote{The fluxes on the image-branes are minus the ones in \eqref{FluxConfig}.}.

In what follows, we will allow only `diagonal' fluxes for the stacks $a,k3,gc$. This means that we will always take $a_i^\sigma=a_i$ $\forall \sigma$, $m_i^\mu=m_i$ $\forall \mu$ and $g_i^\gamma=g_i$ $\forall \gamma$.
On the stack $b$, we will instead allow for a non-diagonal flux, i.e.\ on different branes there can be a different flux.

The gauge flux combines with the pull-back of the bulk $B$-field on the wrapped four-cycle, to give the gauge invariant field strength:
\be
  {\cal F} = F - B\,.
\ee
The gaugino condensation stack needs to be invariant.
For this reason we have to set ${\cal F}_{gc}=0$ and consequently the $B$-field is fixed to be $B = F_{gc}$.
In particular it must be half-integral in the direction of $D_7$.
Let us finally define the following combinations of flux parameters which
will render the subsequent expressions much more compact:
\be
\begin{array}{rclcrcl}\label{FluxCombinations}
 \alpha_a &\equiv& a_2 -g_2 + a_4 -g_4  \,, &\qquad& \beta_a&\equiv& 3a_2 -3g_2 +2a_3 -2g_3 + a_4 -g_4 \,, \\
 \alpha_b^\rho&\equiv& -1-2b_4^\rho+2g_4 \,, &\qquad& \beta_b^\rho&\equiv& -1+b_3^\rho -g_3 -2b_4^\rho+2g_4\,,  \\
  \phi & \equiv & -m_2+g_2-m_4+g_4 \,, &\qquad& \psi&\equiv& m_2 -g_2\,,  \qquad
 \nu\equiv 2 (1 - a_1 + g_1) \,.
\end{array}
\ee
Notice that $\alpha_a+\beta_a$ must be an even number, while $\alpha_b^\rho$ is odd.

There is another source for a FW anomaly, i.e.\ when the pull-back of the NS-NS three-form field strength $H_3$ on the D7-brane world-volume is non-zero. This surely does not  happen for the stacks wrapping the rigid divisors and the K3 fibre, as these cycles have $b_3=0$. For the Whitney brane it is more difficult to compute this number, even if on such a kind of brane $b_3$ is usually zero. Otherwise one has constraints on the possible three-form fluxes that can be switched on.

\subsection{Chiral matter on D7-branes}
\label{ChiralModesSec}

The flux on the D7-branes generates 4D chiral modes.
Apart for the Whitney brane, we always have stacks of $N_A$ branes and $N_A$ image-branes on invariant divisors.
Without fluxes the gauge group is $Sp(2 N_A)$.
When we switch on a flux ${\cal F}_A$ along the diagonal $U(1)$ of the $N_A$ branes,
and a flux $-{\cal F}_A$ on the $N_A$ image-branes, the gauge group gets broken to
$U(N_A) = SU(N_A) \times U(1)$. The $U(1)$ factor
gets an $\mc{O}(M_s)$ mass via the St\"uckelberg mechanism\footnote{The gauge boson eats up the axionic component of the K\"ahler
modulus whose real part is the volume of the divisor dual to the two-cycle supporting the non-vanishing flux.}.
The number of zero-modes in the symmetric and antisymmetric representations of $U(N_A)$ is:
\bea
 I^{(S)}_{A}&=&-\frac12 \int_X [D7_A] \wedge [O7] \wedge {\cal F}_A  -  \int_X [D7_A] \wedge [D7_A] \wedge \mc{F}_A\,, \\
 I^{(A)}_{A}&=&\frac12 \int_X [D7_A] \wedge [O7] \wedge {\cal F}_A  -  \int_X [D7_A] \wedge [D7_A] \wedge \mc{F}_A\,.
\eea
When the stack $A$ with flux ${\cal F}_A$ intersects a stack $B$ with flux ${\cal F}_B$, the number of zero-modes in the bi-fundamental representations
$(N_A,\bar{N}_B)$ and $(N_A,N_B)$ (intersections between branes $A$ and $B$ and between branes $A$ and $B'$) are given by:
\bea
 I_{A\bar{B}}&=& \int_X [D7_A] \wedge [D7_B] \wedge ({\cal F}_A - {\cal F}_B)\,, \\
 I_{AB}&=&\int_X [D7_A] \wedge [D7_B] \wedge ({\cal F}_A + {\cal F}_B)\,.
\eea
If on one $Sp(2N_B)$ stack there is no flux, then $I_{A\bar{B}} = I_{AB}$ and the two $N_A$-representations combine to form an $(N_A,2N_B)$ representation, where now $2N_B$ is the fundamental representation of $Sp(2N_B)$. When a fluxed $Sp(2N)$ stack intersects the Whitney brane, the fields at the intersection are in the $N$ representation of the unbroken $U(N)$ gauge group.

In this paper we will also consider non-diagonal fluxes. On a group of branes of one stack we switch on a flux ${\cal F}_B$ (and $-{\cal F}_B$ on their images) and on the other branes of the stack we take a flux ${\cal F}_{\tilde{B}}$ (and $-{\cal F}_{\tilde{B}}$ on their images). To compute the chiral matter coming from these branes, one can use the above formulae, considering the two groups as different stacks (wrapping the same divisor).

\subsection{D3 tadpole cancellation condition}

For consistency, the total D3-charge must cancel\footnote{The total D5-charge must cancel as well. In our construction this cancellation is automatically implemented by choosing invariant stacks of branes, i.e.\ brane and image-brane wrap the same divisor class. Hence, the D5-charge induced by the flux on one brane is cancelled by the induced charge on the image-brane.}. In our construction we do not want to use D3-branes, as they would introduce new moduli.
Moreover, there are no O3-planes. On the other hand the D7-branes and the O7-plane carry D3-charge. It is given by the
integral over the Calabi-Yau $X$ of minus the six-form in \eqref{GammaD7}, which in this case becomes:
\be
 Q_{(D3)}(D7)
 =  -\frac12 \int_{D7} {\cal F}\wedge {\cal F} - \frac{\chi(D7)}{24} = Q^{\rm flux}_{(D3)}(D7) + Q^{\rm geom}_{(D3)}(D7)\,,
\ee
where we split it, as usual, into flux and geometric D3-charge. The first is positive, while the second is negative. In our configuration, the D7-branes have the same charge as their images. So for the stack $a$, $b$, $k3$ and $gc$ we have just to take:
\be
Q_{(D3)}(D7_i) = 2N_i \left(- \frac12 \int_{D7_i} {\cal F}_i\wedge {\cal F}_i - \frac{\chi(D7_i)}{24}\right)\,,
\ee
where $i=a,b,k3,gc$.
The O7-plane has only negative charge given by:
\be
 Q_{(D3)}(O7) = - \frac{\chi(D7)}{6}\,.
\ee
Another positive contribution to the total D3-charge can come from bulk three-form fluxes:
\be
 Q_{(D3)}(F_3,H_3) \sim \int_X H_3\wedge F_3\,.
\ee

\subsubsection*{D3-charge of the Whitney brane}

To compute the D3-charge of the Whitney brane, we need to use some tricks, as it wraps a singular four-cycle.
The Whitney brane on the divisor $D_W=2D_P$ can be understood as the recombination of a brane on $D_P$ with its image in the same class. In the recombination,
the total charge is conserved (as a consequence, ${\cal F}$ is trivial as a two-form on $X$). Hence, we can compute the D3-charge of a Whitney brane, by computing it in the (non-singular) split case. Doing this, we get:
\be
Q_{(D3)}(D7_W) =  -  \int_X \left( D_P\wedge \left(\tfrac12 D_P -B - S\right)^2  + \tfrac{1}{12}\,D_P\wedge c_2(D_P)\right)\,,
\ee
where $S$ is an arbitrary integral class on $X$, that does not contribute to the chiral intersections. When the Whitney brane is split, the flux is $F_P = \frac12 D_P - S$.

There are conditions on the divisors $D_P$ and $S$. If they are satisfied, the Whitney brane is consistent
with supersymmetry (holomorphicity) and it is stable (it is not forced to split into a brane/image-brane pair) \cite{Collinucci:2008pf}:
\be
D_P>0, \qquad D_P-[O7]>0, \qquad \frac{[O7]}{2}\leq S+B \leq D_P - \frac{[O7]}{2}\,.
\label{ConstrWIIN}
\ee
In particular, the last condition gives the range on which the flux can vary:
\be
  \left|{\cal F}_P\right| = \left| \frac{D_P}{2} - S - B \right| \leq \left|\frac{[O7]}{2} - \frac{D_P}{2} \right|
\ee
We can parameterise such a flux as follows:
\be
 {\cal F}_P = \frac12 \left(N_{gc}+N_{k3}-2k-1\right)D_7+\left(\frac{N_{k3}}{2}-q\right)D_6 + \left(\frac{N_{b}}{2}-k_b\right)D_5+\left(\frac{N_{a}}{2}-k_a\right)D_4\,,\nonumber
\ee
with the integers $k,q,k_a,k_b$  chosen such that \eqref{ConstrWIIN} are satisfied\footnote{For example if $N_{k3}=0$ and $N_{gc}$ is odd,
then the coefficient of $D_7$ can be cancelled by a choice of $k$: the coefficient of $D_P$ along $D_7$ is odd and the $D_P/2$ term in the flux
can be cancelled by the half-integer $B$-field (recall that we took $B=F_{gc}$ that is half-integer in the direction of $D_7$).}.

\subsection{K-theory constraints}

In order to have a fully consistent brane set-up, the cancellation of the tadpoles is not enough. One has to check that all K-theoretic charges sum up to zero \cite{WittenKth,WitMooreKth}. This is not a simple task when one considers the torsion charges. Fortunately, there is a probe brane argument~\cite{Uranga} that gives us an equivalent condition. In the absence of a global $SU(2)$ Witten anomaly on any probe brane wrapping an invariant four-cycle transverse to the O7-plane and supporting symplectic Chan-Paton factors, the K-theory torsion charges are cancelled. Hence, a D7-brane configuration passes the check, if for any such cycle the wrapped probe brane has an even number of fundamental chiral $Sp$-representations.

In our constructions, all divisor classes are invariant under the orientifold involution. Hence, we will check this constraint for any divisor whose representatives are transverse to the O7-plane. If the four-cycle wrapped by the probe brane is \textit{non-Spin}, FW anomaly cancellation will imply the existence of a flux that breaks the group to a non-symplectic one, avoiding the possible anomaly. If however the flux on the probe brane can be set to zero, one has to compute the chiral intersections with all the D7-branes in the given configuration and find an even number of chiral states.

\section{Example with two D-term conditions}
\label{Ex1}

In this first example, we wrap no brane around the K3 divisor (i.e.\ $N_{k3}=0$), while we choose the following values for the other $N_i$:
\be
 N_a = 5\,, \qquad N_b=2\,, \qquad N_{gc} = 4 \,.
\ee
These numbers satisfy the conditions \eqref{CondNiFact}.
Moreover, we switch on non-zero fluxes ${\cal F}_a$ and ${\cal F}_b$ on the stacks on $D_4$ and $D_5$, such that the gauge group is:
\be
  U(5) \times U(1) \times U(1)\times Sp(8) \rightarrow SU(5) \times U(1) \times Sp(8) \,.
\ee
The last factor is relative to the brane wrapping the `diagonal' del Pezzo divisor supporting gaugino condensation.
The first two factors are relative to the visible sector which
lives on rigid divisors with $h^{1,0}=0$ (so that we avoid zero-modes in the adjoint representation). On the stack wrapping $D_4$ we have switched on a diagonal flux breaking $Sp(10)$ to $U(5)$ (i.e.\ ${\cal F}_a^\sigma={\cal F}_a$ $\forall \sigma$), while on the stack wrapping $D_5$ we have taken a non-diagonal flux breaking $Sp(4)$ to $U(1)\times U(1)$ (i.e.\ we take ${\cal F}_b^{\rho=1}\not=\pm{\cal F}_b^{\rho=2}$).
For each stack, the $U(1)$ associated with the Cartan along which we switch on the flux gets a St\"uckelberg mass of $\mc{O}(M_s)$.
Thus we obtain an $SU(5)$ GUT-like theory plus a massless $U(1)$. However, as we shall see in section \ref{Fpot1st},
this massless $U(1)$ is much more weakly coupled than $SU(5)$ at the string scale, and so it behaves somehow as a \textit{dark force}
with interesting phenomenological applications \cite{HiddenPhotons}.
If an $SU(5)$ singlet gets a VEV, this Abelian gauge boson could become massive by eating up the corresponding phase.
We stress that we did not look for this extra light $U(1)$ but we have been forced to introduce it by consistency constraints (as we will explain later, cancellation of K-theory torsion charges forces $N_b$ to be even).

Given that we want only one D-term condition coming from the stack $b$, we choose the flux so that ${\cal F}_b^{\rho=2}$ and ${\cal F}_b^{\rho=1}$ are proportional. In particular we take:
\be
 b_i^{\rho=1}\equiv b_i\,, \qquad \qquad b_i^{\rho=2}=3b_i -2g_i +\delta_i^4\,.
\ee
Consequently, we have $\alpha_b^{\rho=2}=3\alpha_b^{\rho=1}\equiv 3\alpha_b$ and $\beta_b^{\rho=2}=3\beta_b^{\rho=1}\equiv 3\beta_b$.

The D7 tadpole cancellation condition is saturated by the Whitney brane $W=0$, whose homology class is:
\bea
 [D7_W] &=& 8 [O7] - 2N_a [D7_a]- 2N_b [D7_b] - 2N_{gc} [D7_{gc}] \nonumber\\
          &=& 2\left(7\,\G_2-7\,\G_1 + 5\,\G_3 - \,\G_4 \right).
\eea
The brane configuration is summarised in the following table:
\be
 \begin{array}{c|c|c|c|c}
   \mbox{D7-stack} & D7_a & D7_b &  D7_{gc} & D7_W\\ \hline
   N_i & 5 & 2 &  4 & - \\ \hline
   \mbox{divisor class} & D_4 & D_5 & D_7 & [D7_W]\\ \hline
   \mbox{topology} & \mbox{rigid} & \mbox{rigid} & {\rm dP}_7 & \mbox{Whitney brane}\\
 \end{array}
\ee

\subsection{Chiral intersections}

Using the formulae in section \ref{ChiralModesSec} the numbers of chiral zero-modes are given by:
\be
\begin{array}{rclcrcl}
 I^{(A)}_a &=& 2\beta_a-\nu\,, &\qquad& I^{(S)}_a &=& -2\beta_a+3\nu\,, \\
 I^{(S)}_{b^1} &=& 0\,,  &\qquad&  I^{(S)}_{b^2} &=& 0 \,, \\
 I_{b^2\bar{b}^1} &=& -4\beta_b\,,  &\qquad&  I_{b^2b^1} &=& -8\beta_b \,, \\
 I_{a\bar{b}^1} &=& -3\alpha_a+\beta_a+\alpha_b-2\beta_b\,, &\qquad& I_{ab^1} &=& -3\alpha_a+\beta_a-\alpha_b+2\beta_b\,, \\
 I_{a\bar{b}^2} &=& -3\alpha_a+\beta_a+3\alpha_b-6\beta_b\,, &\qquad& I_{ab^2} &=& -3\alpha_a+\beta_a-3\alpha_b+6\beta_b\,, \\
 I_{aW} &=& 12(\alpha_a+\beta_a)+ 14\nu\,, &\qquad& I_{b^1W} &=& 5(2\alpha_b+2\beta_b)\,,    \\
 I_{b^2W} &=& 15(2\alpha_b+2\beta_b) &\qquad& I_{agc}  &=& \nu\,,
 \end{array}
\ee
where these expressions are given in terms of the flux-dependent parameters defined in \eqref{FluxCombinations}.
The only non-zero chiral intersection of the gaugino condensation stack is $I_{agc}$.
We set it to zero by choosing the fluxes such that $\nu=0$, i.e.:
\be\label{chIntgc}
\nu = 0 \qquad \Rightarrow \qquad a_1 = g_1 - 1 \,.
\ee

\subsection{D3 charge}

The D3-charge of the flux on the Whitney brane is minimal when
$k=6$, $q=4$, $k_a=4$ and $k_b=3$. With these values, the total D3-charge is given by:
\bea
 Q_{(D3)}^{\rm tot} &=&  -438 + 20 \alpha_a (2\alpha_a -\beta_a) +  10\alpha_b( \alpha_b - 2 \beta_b)\,.
\eea
If $Q_{(D3)}^{\rm tot}<0$, there is space to
switch on three-form fluxes in the bulk and two-form fluxes on the Whitney brane,
in order to fix the complex structure and the D7-brane moduli.

\subsection{K-theory charge}

As we have said, in order for the background to be consistent, it is necessary to cancel also the K-theoretic torsion charges. We check this using the probe brane argument described above. We wrap a probe D7-brane on each invariant divisor and, if we can have zero flux on it (in order to have an $Sp$-group), we have to verify that the number of its chiral intersections is even. In our configuration, an invariant cycle on which we can have non-zero flux\footnote{Remind that the non-integral part of the $B$-field is fixed by requiring ${\cal F}_{gc}=0$.} is the K3 fibre $D_1$. A problem may occur on the intersection with $D_5$. If we had $N_b=1$, then the number of chiral modes at the intersection with $D_1$ would be $1 + 2(b_4-g_4)$, i.e.\ an odd number. This is the reason why we have chosen $N_b=2$; combined with our choice of fluxes (${\cal F}_b^{\rho=2}=3{\cal F}_b^{\rho=1}$), it gives an even number of fundamentals for a probe brane wrapping $D_1$. We have checked that this is also true for all probe branes wrapping the other invariant divisors (on which we can have zero flux).  From explicit computations one also sees that any probe brane wrapping an invariant divisor has always even chiral intersection numbers with branes wrapping the divisor $D_4$ (for any value of $N_a$).

\subsection{A consistent flux choice}

The following choice of flux numbers is consistent with all our requirements:
\be\label{FluxChoice1stModel}
\alpha_a=1, \qquad \beta_a=-1, \qquad \alpha_b=9, \qquad \beta_b=4, \qquad \nu=0\,.
\ee
The total D3-charge is $Q_{(D3)}^{\rm tot} = -318$ while the chiral intersections are:
\be
\begin{array}{lclclcl}
I^{(A)}_a = -2\,, &\qquad & I^{(S)}_a = 2\,, &\qquad &
I^{(S)}_{b^1} = 0\,,  &\qquad &  I^{(S)}_{b^2} = 0 \,, \\
I_{b^2\bar{b}^1} = -16\,,  &\qquad &  I_{b^2b^1} = -32 \,,&\qquad &
I_{a\bar{b}^1} = -3\,, &\qquad & I_{ab^1} = -5\,, \\
I_{a\bar{b}^2} = -1 \,, &\qquad & I_{ab^2} = -7\,, &\qquad &
I_{aW} = 0\,, &\qquad &  \\
I_{b^1W} = 106 &\qquad & I_{b^2W} = 318 &\qquad & I_{agc}  = 0\,.
\end{array}
\ee
The modes in the ${\bf \bar{5}}$ representation of $SU(5)$ are given by $ I_{a\bar{b}}+ I_{ab}+I_{aW}$.
We also have two modes in the ${\bf 10}$ and two in the ${\bf 15}$ representations of $SU(5)$.
Finally we have a large number of $SU(5)$-singlets.

\subsection{K\"ahler moduli stabilisation}

\subsubsection{D-term potential}

For generic fluxes (satisfying \eqref{chIntgc}) and with the given choice for the $B$-field,
the only non-trivial FI-terms are:
\bea
\label{xiabk3}
 \xi_a &=& \frac{1}{4\pi\vo}\int_X [D7_a]\wedge J\wedge {\cal F}_a =\frac{1}{4\pi\vo}\left[ (\beta_a-\alpha_a)(\ti_1+\ti_2) + 2\alpha_a\ti_3\right]\,, \\
 \xi_b &=& \frac{1}{4\pi\vo}\int_X [D7_b]\wedge J\wedge {\cal F}_b= \frac{1}{4\pi\vo}\left[\alpha_b\,\ti_3 -(\alpha_b -2\beta_b) \ti_4\right]\,.
\eea
The second one is the FI-terms coming from the brane $\rho=1$ of the $D7_b$ stack. The one relative to $\rho=2$ is proportional to this, and so it gives the same D-term condition.

The system of equations $(\xi_a,\xi_b)=(0,0)$, has the following solution:
\be
 \ti_3 = \frac{\alpha_a - \beta_a}{2\alpha_a} \,(\ti_1+ \ti_2) \,,\qquad
 \ti_4 =  \frac{\alpha_b(\alpha_a-\beta_a)}{2\alpha_a(\alpha_b-2\beta_b)}\, (\ti_1+ \ti_2) \,,
\ee
which, expressed in terms of the original $t_i$ variables, takes the form:
\be
 \ri_2 = \frac{\alpha_a(4\beta_b-3\alpha_b)+\beta_a\alpha_b}{\alpha_b(\alpha_a-\beta_a)}\, \ri_4 \,,\qquad\qquad
 \ri_3 = \frac{2\beta_b}{\alpha_b}\, \ri_4 \,.
\label{Dt1st}
\ee
The flux parameters must be chosen to ensure that these solutions are inside the K\"ahler cone.
For the choice taken in \eqref{FluxChoice1stModel}, (\ref{Dt1st}) simplifies to:
\be
 \ri_2= -\frac{10}{9}\,\ri_4\,, \qquad\qquad \ri_3=\frac{8}{9}\,\ri_4\,.
\label{Dcond1st}
\ee
Substituting this result in the volumes of the relevant divisors, we obtain:
\be
 \tau_4 = \frac{1}{27} \,\ri_4^2 -\ri_1^2\,,\qquad
 \tau_5 =  \frac 29\,\ri_4^2\,, \qquad \vo =  \frac 13 \left[ \frac{86}{729} (- \ri_4)^3 +\ri_1^3\right]\,.
\ee

\subsubsection{F-term potential}
\label{Fpot1st}

The two D-term conditions (\ref{Dcond1st}) fix two K\"ahler moduli at leading order. These moduli
can be parameterised as the volumes of the divisors supporting the D7-branes on whose world-volume we have turned on a flux:
$\tau_4$ and $\tau_5$. These two moduli are fixed at:
\be
\tau_4=\frac{3}{19}\,\tau_1-\tau_7\,, \qquad \tau_5=\frac{18}{19}\,\tau_1\,,
\ee
and disappear from the effective field theory since they acquire a mass of the order the string scale. In addition the corresponding
axions get eaten up by the $U(1)$s which get massive via the St\"uckelberg mechanism.

We can then study the effective field theory in terms of the two remaining K\"ahler moduli $\tau_1$ and $\tau_7$.
The D-term stabilisation (\ref{Dcond1st}) gives a volume of the form:
\be
\vo=\alpha\left(\tau_1^{3/2}-\gamma\tau_7^{3/2}\right)\,,
\ee
where $\alpha= 86/(57\sqrt{19})$, $\gamma= 1/(3\alpha)$, and:
\be
\tau_1= \frac{19}{81}\, t_4^2\,, \qquad \tau_7=t_1^2\,.
\ee
We shall now show that these two remaining moduli can be fixed within the K\"ahler cone and the
regime of validity of the effective field theory by the interplay of $\alpha'$ and non-perturbative corrections.

Writing the K\"ahler moduli as $T_i=\tau_i + i \zeta_i$, the K\"ahler potential and superpotential
of the low-energy 4D $\mc{N}=1$ supergravity read (for $N_{gc}=4$):
\be
K=-2\ln\left(\vo+\frac{\hat\xi}{2}\right)\,,\qquad W=W_0 + A \,e^{-\frac{2\pi T_7}{N_{gc}+1}}=W_0+ A \,e^{-\frac{2\pi T_7}{5}}\,,
\ee
where $\hat\xi=\xi/g_s^{3/2}$ with $\xi\simeq 0.5$.
The form of the $\mc{N}=1$ F-term supergravity scalar potential is
(after $T_7$-axion minimisation which fixes the sign of the second term as negative):
\be
V=\frac{1}{(\vo+\frac{\hat\xi}{2})^2}\left[\frac{4\pi ^2 A^2}{25} \,K^{7\bar{7}}\, e^{-\frac{4\pi  \tau_7}{5}}
- \frac{4\pi}{5}\,|K^{7\bar{j}}K_j \,A  W_0| \,e^{-\frac{2\pi  \tau_7}{5}}+\frac{3 W_0^2 \hat{\xi} \left(\vo^2+7 \vo \hat\xi+\hat\xi^2\right)}{(\vo-\hat\xi) (2 \vo+\hat\xi)^2} \right],
\label{PotN}
\ee
where the $\alpha'$ corrected $7\bar{7}$ element of the inverse K\"ahler metric is given by:
\be
K^{7\bar{7}}= \frac{8 \sqrt{\tau_7}\, \vo\left(1+ \frac{\tau_7^{3/2}}{2\vo}\right) \left(1+\frac{\hat\xi}{2 \vo}\right)
\left(1-\frac{\hat\xi}{4\vo+2 \tau_7^{3/2}}\right)}{\left(1-\frac{\hat\xi}{4\vo}\right)}\,,
\label{InvKN}
\ee
and:
\be
|K^{7\bar{j}}K_j| = 4 \tau_7 \left[\frac{3}{2 \left(1-\frac{\hat\xi}{4 \vo}\right)}-1\right]\,.
\ee
Given that the combination of the D-term constraints (\ref{Dcond1st}) and the requirement of
obtaining a small value of the visible sector gauge coupling $\alpha_{\rm vis}^{-1}\simeq\tau_4$
does not allow us to perform a limit where some divisors become exponentially large while others stay relatively small,
we cannot approximate $K^{7\bar{7}}$ taking $1+ \frac{\tau_7^{3/2}}{2\vo}= 1+\epsilon$ with $\epsilon\ll 1$.
Therefore, we shall keep this term showing however that $\epsilon \simeq 0.008$.
Moreover, in order to trust the stability of our vacuum, we need also to make sure that we have control over the perturbative expansion,
in the sense that the $\alpha'$ corrections, controlled by the parameter $\hat\xi/\vo$, should be small.
We will indeed show that $\hat\xi/\vo\simeq 0.01$ justifying the neglecting of higher order $\alpha'$ and perturbative corrections (like string loops which
are subleading due to the no-scale structure). More precisely, these corrections might slightly modify the exact position of the minimum
but not the stability and the main phenomenological features of the vacuum.

The general expression for the scalar potential (\ref{PotN}) is quite complicated and so its vacuum structure cannot be studied analytically.
Therefore, we will perform a numerical study to show the presence of a global and stable minimum.
However we shall first analyse our potential analytically in the approximation of small $\tau_7$ and $\alpha'$ corrections.
Due to the good agreement between this analytic study and the full numerical minimisation, we shall understand
the qualitative behaviour of the potential and trust our approximations.

The approximated potential looks like (taking both $A$ and $W_0$ positive):
\be
V\simeq \frac{32}{25}\,\pi ^2 A^2 \frac{\sqrt{\tau_7}}{\vo} \left(1+ \frac{\tau_7^{3/2}}{2\vo}\right)\, e^{-\frac{4\pi  \tau_7}{5}}
- \frac{8}{5}\,\pi A  W_0 \,\frac{\tau_7}{\vo^2}\,e^{-\frac{2\pi  \tau_7}{5}}+\frac{3 W_0^2 \hat\xi }{4 \vo^3}\left(1+\frac{7 \hat\xi}{\vo}\right)\,,
\label{PotApprN}
\ee
where we have neglected the $\alpha'$ corrections in both $K^{7\bar{7}}$ and $|K^{7\bar{j}}K_j|$ but
we have kept the leading order one in the third term in (\ref{PotApprN}) due to the large coefficient $7$.
The two minimisation conditions $\partial V/\partial \tau_7=0$ and $\partial V/\partial \vo=0$ in the limit $2\pi\tau_7/5\gg 1$ which
guarantees the neglecting of higher order non-perturbative effects, simplify to:
\be
\frac{\partial V}{\partial \tau_7}=0 \quad \Leftrightarrow \quad
\vo\simeq\frac{5 W_0 \,\sqrt{\tau_7}}{8 \pi  A}\left(1-\frac{\tau_7^{3/2}}{2 \vo}\right)\,e^{\frac{2\pi  \tau_7}{5}}\,,
\label{firstN}
\ee
and:
\be
\frac{\partial V}{\partial \vo}=0 \quad \Leftrightarrow \quad
a \vo^2 - b \vo +c =0\,,
\label{secondN}
\ee
where:
\be
a = 128 \pi ^2 A^2 \sqrt{\tau_7} \left(1+\frac{\tau_7^{3/2}}{\vo}\right), \quad
b= 320 \pi  \tau_7 \,A W_0\,e^{\frac{2\pi  \tau_7}{5}}, \quad
c= 225 W_0^2 \hat\xi\left(1+\frac{28 \hat\xi}{3\vo}\right)\,  e^{\frac{4\pi  \tau_7}{5}}\,. \nonumber
\ee
The solution of (\ref{secondN}) gives:
\be
\vo= \left[1\pm\sqrt{1-\frac{9 \hat\xi}{8
   \tau_7^{3/2}}\left(1+\frac{28 \hat\xi}{3\vo}\right) \left(1+\frac{\tau_7^{3/2}}{\vo}\right)}\right]\frac{5 W_0\,\sqrt{\tau_7} }{4\pi  A }\left(1-\frac{\tau_7^{3/2}}{\vo}\right)\,e^{\frac{2\pi \tau_7}{5}}\,.
   \label{thirdNn}
\ee
Combining (\ref{firstN}) with (\ref{thirdNn}) we realise that we have to focus on the solution with the minus sign obtaining:
\be
1-\frac 12 \left(1+\frac{\tau_7^{3/2}}{2 \vo}\right)
\simeq \sqrt{1-\frac{9 \hat\xi}{8
   \tau_7^{3/2}}\left(1+\frac{28 \hat\xi}{3\vo}\right) \left(1+\frac{\tau_7^{3/2}}{\vo}\right)}\,.
\ee
Taking the square, the previous expression reduces to:
\be
\tau_7 \simeq
\left[\frac{3 \xi}{2}\left(1+\frac{28\hat\xi}{3\vo}\right)\right]^{2/3}
\left(1+\frac{4\tau_7^{3/2}}{9\vo}\right) \frac{1}{g_s}\,.
\label{finalN}
\ee
Combining (\ref{firstN}) and (\ref{finalN}) we notice that at the minimum
$\tau_7$ and $\vo$ scale as
\be
\tau_7 \sim g_s^{-1} \quad \text{and} \quad
\vo\sim W_0 \,e^{\frac{2\pi  \tau_7}{5}} \sim W_0 \,e^{\frac{2\pi}{5}\frac{1}{g_s}}\,,
\ee
and so we realise that the only way to get a minimum within the K\"ahler cone with all the moduli
of the same order of magnitude, as required by the D-term conditions (\ref{Dcond1st}), is to fine tune $W_0\ll 1$.
This tuning prevents us to build a standard LVS but, besides fixing all the moduli,
opens up the possibility to obtain both GUT theories and TeV-scale SUSY.
In fact, as we shall see below, in order to get a phenomenologically
viable GUT scale of the order $M_{\rm GUT}\sim 10^{16}$ GeV we need to choose $\vo \sim \mc{O}(10^3)$.
In turn, the gravitino mass (and the soft terms
generated via gravity mediation) turns out to be of the order the TeV scale only if we choose $W_0 \sim \mc{O}(10^{-10})$.

\begin{figure}
\begin{center}
\scalebox{.6}{\includegraphics*{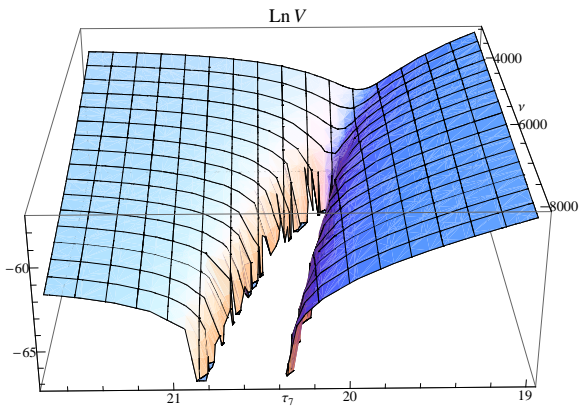}}
\caption{Plot of $\ln(V)$ as a function of $\vo$ and $\tau_7$. In the region where the logarithm is
 undefined the potential becomes negative and develops a global minimum whose presence is guaranteed by the
 fact that $V\to 0$ at infinite volume.}\label{fig1}
 \end{center}
\end{figure}

Let us now present a choice of the underlying parameters that gives rise to a global minimum
with the required phenomenological features:
\be
W_0\simeq 5.51\cdot 10^{-9}\,, \qquad A=0.10\,, \qquad g_s\simeq 0.04\,.
\ee
We find numerically that (\ref{firstN}) and (\ref{finalN}) have a solution at:
\be
\langle\tau_7\rangle_{\rm app}\simeq 20.25\,,   \qquad \langle\vo\rangle_{\rm app} \simeq 5507.23\,,
\ee
which is in good agreement with the result obtained by minimising the whole potential (\ref{PotN}) which
looks like:
\be
\langle\tau_7\rangle\simeq 20.30\,,   \qquad \langle\vo\rangle \simeq 5732.80\,.
\ee
This qualitative agreement justifies the validity of our approximations.
The potential for $\vo$ and $\tau_7$ is plotted in Fig. \ref{fig1}.

Due to the presence of $\alpha'$ corrections, this minimum breaks supersymmetry spontaneously inducing
non-zero F-terms of the K\"ahler moduli. Moreover, the minimum is AdS with a depth of the order
(introducing the correct prefactor $g_s e^{K_{cs}}/(8\pi)$ and setting the VEV of the K\"ahler potential
for the complex structure moduli such that $e^{K_{cs}}=1$):
\be
\langle V\rangle \simeq - 4.87\cdot 10^{-31} M_P^4\,,
\ee
and so an up-lifting mechanism is needed. This can be done using one of the various
methods proposed in the literature (D-terms from magnetised D7-branes \cite{bkq},
F-terms from a hidden sector \cite{ss}, inclusion of $\overline{D3}$
branes \cite{kklt}, etc.).

Moreover, we obtain nice phenomenological scales since the string scale is of the order:
\be
M_s \simeq \frac{M_P}{\sqrt{4 \pi \vo}}\simeq 8.94\cdot 10^{15} \,{\rm GeV},
\ee
allowing GUT model building since the GUT scale is given by:
\be
 M_{\rm GUT} \simeq M_s \,\vo^{1/6}\simeq 3.78\cdot 10^{16} \,{\rm GeV}.
\ee
The flux-corrected value of the GUT coupling turns out to be:
\be
\alpha_{\rm GUT}^{-1}= \tau_4+ \frac{1}{2 g_s}\int_{D_4}\mc{F}_4\wedge \mc{F}_4\simeq 150\,.
\label{alphaGUT}
\ee
Notice that we can reproduce the right order of magnitude but not the exact phenomenological value
$\alpha_{\rm GUT}^{-1}\simeq 25$ because the requirement of getting the right GUT-scale and
TeV-scale SUSY fixes the values of $\vo$ and $W_0$. These, in turn, fix the value of $\tau_7=t_1^2\sim \mc{O}(20)$
which then, using the K\"ahler cone condition $r_2>0$ with the D-term constraint $t_2=-10\,t_4/9$,
sets a lower bound on $\tau_4>\mc{O}(40)$.
Nevertheless we do not have the right matter spectrum and so we do not expect to
reproduce the standard picture for gauge coupling unification.

Notice also that the coupling (\ref{alphaGUT}) is much stronger than the coupling of the
massless $U(1)$ left over on $D_5$. In fact, after performing the proper diagonalisation
to identify the massless Abelian gauge boson on $D_5$,
this coupling turns out to be of the order $\alpha_{\rm dark}^{-1}\simeq 1206$.
As we have already commented at the beginning of section \ref{Ex1}, this weakly interacting
extra $U(1)$ might behave as a \textit{dark force} with interesting phenomenological applications \cite{HiddenPhotons}.

TeV-scale supersymmetry is obtained because the gravitino mass turns out to be:
\be
m_{3/2}= e^{K/2} W_0 M_P = \sqrt{\frac{g_s}{8\pi}} \frac{W_0 M_P}{\vo} \simeq 95.63 \,{\rm TeV}\,,
\ee
and gravity mediation would yield \cite{SUSYbreaking}:
\be
M_{\rm soft}\simeq \frac{m_{3/2}}{\ln\left(M_P/m_{3/2}\right)}\simeq 3.1 \,{\rm TeV}\,.
\ee
We can also check that all the K\"ahler moduli are fixed within the K\"ahler cone since:
\be
r_1 = 4.5 >0\,, \qquad r_2\simeq 1.28>0\,, \qquad r_3\simeq 5.78 >0\,, \qquad r_4\simeq 52.03 >0\,. \notag
\ee
Moreover, the volumes of all the divisors are fixed larger than unity, and so above the string scale
where we can trust the effective field theory since\footnote{Also all the other divisors have large volumes.}:
\be
\langle\tau_1\rangle \simeq 634.92\,, \qquad \langle\tau_4\rangle = 80\,,\qquad \langle\tau_5\rangle\simeq 601.50\,,
\qquad \langle\tau_7\rangle = 20.25\,.
\ee
The values of the stabilised dual two-cycle volumes are also larger than unity which is a necessary condition
to neglect $g_s$ corrections to the K\"ahler potential. In fact, if we consider
the basis formed by all the wrapped divisors $\{D_1,D_4,D_5,D_7\}$ defined as:
\be
\G_1 = D_7,\qquad \G_2=D_1+D_4+D_5+D_7, \qquad \G_3=D_1, \qquad \G_4=D_5\,,
\ee
and expand the K\"ahler form as $J=\sum_i x_i D_i$, we find that:
\be
x_1=\ri_2+\ri_3\simeq 11.56\,,\quad x_4=\ri_2\simeq 57.81\,,
\quad x_5=\ri_2+\ri_4\simeq 5.78\,, \quad x_7 = \ri_1 + \ri_2\simeq 53.31\,. \nonumber
\ee
We finally point out that the largeness of the volume justifies the neglecting of higher $\alpha'$
corrections and a relatively small $\tau_7$. In fact, the expansion parameter is $\hat\xi/\langle\vo\rangle\sim 0.01$ which is small enough to consider just
the leading order $\alpha'$ correction, and $\langle\tau_7\rangle^{3/2}/(2\langle\vo\rangle)\sim 0.008$.

\section{Example with one D-term condition}
\label{Ex2}

In this example we will allow only one non-trivial D-term condition, instead of two, and we will fix the otherwise
unfixed combination of the K\"ahler moduli, by string loop corrections to the K\"ahler potential.
Given that the D-term condition will leave two flat directions and
the visible gauge coupling will constraint the size of just one of them, we shall be
able to stabilise the moduli such that the other direction is exponentially large. This is the first
realisation of a LVS in a globally consistent model.

We shall focus on the same Calabi-Yau three-fold as in the previous example,
but without wrapping any brane on $D_5$ (i.e.\ $N_b=0$).
Instead of that, we will wrap the K3 divisor. The number of branes in each stack is:
\be
 N_a=3\,, \qquad N_{k3}=1\,, \qquad N_{gc}=3 \,.
\ee
We switch on a non-zero flux ${\cal F}_a$ on $D7_a$, while we set the flux on K3 to be zero\footnote{The Freed-Witten anomaly cancellation does not force any flux as $c_1(K3)=0$}. This is possible since the non-integral part of $B$ pulled back to $D_1$ is zero.
In this case the $Sp(2)\cong SU(2)$ group is unbroken and we do not have a D-term condition coming from this stack.
The gauge group is:
\be
 U(3) \times SU(2) \times Sp(6)\rightarrow  SU(3) \times SU(2) \times Sp(6)\,,
\ee
where the first two factors are relative to the visible sector.
The D7 tadpole cancellation condition is saturated by the Whitney brane $W=0$ whose homology class is now:
\bea
 [D7_W] &=& 8 [O7] - 2N_a [D7_a]- 2N_{k3} [D7_{k3}] - 2N_{gc} [D7_{gc}] \nonumber\\
          &=& 2\left(-8\,\G_1 +9\,\G_2 + 2\,\G_3 - \G_4\right)\,.
\eea
The brane configuration is summarised in the following table:
\be
 \begin{array}{c|c|c|c|c}
   \mbox{D7-stack} & D7_a & D7_{k3} &  D7_{gc} & D7_W\\ \hline
      N_i & 3 & 1 & 3 & - \\ \hline
   \mbox{divisor class} & D_4 & D_1 & D_7 & [D7_W]\\ \hline
   \mbox{topology} & \mbox{rigid} & {\rm K3}  & {\rm dP}_7 & \mbox{Whitney brane}\\
 \end{array}
\ee

\subsection{Chiral Intersections}

Using the rules reviewed in section \ref{ChiralModesSec}, we compute
the chiral intersections:
\be
\begin{array}{rclcrcl}
 I^{(A)}_a &=& 2\beta_a-\nu\,, &\qquad& I^{(S)}_a &=& -2\beta_a+3\nu\,, \\
 I_{ak3} &=& 2\alpha_a\,, &\qquad & I_{aW} &=& 4(4\beta_a-\alpha_a) + 8\,\nu \\  I_{k3W} &=& 0\, &\qquad& I_{agc}  &=& \nu\,,
 \end{array}
\ee
which are functions of the combinations \eqref{FluxCombinations}.
Again, the only non-zero chiral intersection of the gaugino condensation stack is $I_{agc}$, that we set to zero by choosing $\nu=0$.

\subsection{D3 charge}

The D3-charge of the Whitney brane is minimal when $k=6$, $q=4$, $k_a=4$ and $k_b=3$. In this case, the total D3-charge, after setting $\nu=0$, turns out to be:
\bea
 Q_{(D3)}^{\rm tot} &=& -624 + 6\alpha_a(2\alpha_a - \beta_a) \,.
\eea
If this total charge is negative, there is space to
switch on three-form fluxes in order to fix the complex structure moduli and the D7-brane moduli of the Whitney brane.

\subsection{K-theory charge}

In this example, the only fluxed branes are the ones wrapping the divisor $D_4$. Luckily, this divisor has even chiral intersections with any other invariant divisor on which we can have a vanishing flux. Therefore, any probe brane with $Sp$-gauge group is free from Witten anomaly. By the probe brane argument of \cite{Uranga} we conclude that the K-theoretic torsion charges are cancelled.

\subsection{A consistent flux choice}

The following choice of flux numbers is consistent with all requirements:
\be
\alpha_a=1\,, \qquad \beta_a=-1\,, \qquad \nu=0\,.
\ee
The total D3-charge is $Q_{(D3)}^{\rm tot} = -606$ while the non-zero chiral intersections look like:
\be
 I^{(A)}_a = -2\,, \qquad I^{(S)}_a = 2\,, \qquad I_{ak3} = 2\,, \qquad I_{aW} =  -20\,.
\ee

\subsection{K\"ahler moduli stabilisation}

\subsubsection{D-term potential}

When we switch on a flux only along the $D7_a$ stack (with $\nu=0$), there is only one non-trivial FI-term:
\bea
\label{xiabk3II}
 \xi_a &=& \frac{1}{4\pi\vo}\int_X [D7_a] \wedge J\wedge {\cal F}_a
 =\frac{1}{4\pi\vo}\left[ (\beta_a-\alpha_a)(\ti_1+\ti_2) + 2\alpha_a\ti_3\right]\,.
\eea
The solution to the D-term condition $\xi_a=0$ is:
\be
 \ti_3 =  \left(1-\frac{\beta_a}{\alpha_a}\right)\frac{ \ti_1+ \ti_2 }{2}\,,
\ee
which, in terms of the variables $\ri_i$, becomes:
\be
\ri_3 = \left(1-\frac{\beta_a}{\alpha_a}\right)\frac{\ri_2}{2} + \left(3-\frac{\beta_a}{\alpha_a}\right)\frac{\ri_4}{2}\,.
\ee
With the chosen fluxes we have:
\be
\ti_3=\ti_1+\ti_2\,, \qquad \text{or} \qquad \ri_3 = \ri_2+2\ri_4 \,.
\label{Dcond2nd}
\ee
Substituting this result in the volumes of the relevant divisors, we obtain:
\bea
\tau_1 &=& (\ri_2 + \ri_4) (\ri_2-\ri_4)\,, \qquad
\tau_4 = 3 (\ri_2 + \ri_4)^2 - \ri_1^2\,, \notag \\
&& \vo = \frac 13 \left[ (5 \ri_2 - 4 \ri_4) (\ri_2 + \ri_4)^2\right]+\frac 13 \,\ri_1^3\,.
\label{AnisVol}
\eea

\subsubsection{F-term potential}

The D-term (\ref{Dcond2nd}) fixes the K\"ahler modulus $\tau_4$ supporting
the visible sector D7-branes on whose world-volume we turn on a flux. This modulus is fixed at:
\be
\tau_4=3\left(\tau_1-\tau_5\right)-\tau_7\,,
\ee
and disappears from the low-energy theory since it acquires an $\mc{O}(M_s)$ mass. The corresponding
axion gets eaten up by the $U(1)$ which gets also an $\mc{O}(M_s)$ St\"uckelberg mass.

The volume (\ref{AnisVol}) can then be written in terms of the three remaining K\"ahler moduli $\tau_1$, $\tau_5$ and $\tau_7$
as:
\be
\vo=\frac 16 \sqrt{\tau_1 - \tau_5} \left(10 \tau_1 - \tau_5\right) - \frac 13 \,\tau_7^{3/2}\,.
\label{overallVol}
\ee
This expression can be simplified by noting that in order to get a visible sector
gauge coupling $\alpha_{\rm vis}^{-1}=\tau_4$ which is not too small, the combination $\left(\tau_1-\tau_5\right)$
has to be fixed small. Hence, performing the following change of variables:
\be
\tau_s\equiv\tau_1-\tau_5=\left(\ri_2+\ri_4\right)^2, \qquad \tau_b\equiv\frac{10\tau_1-\tau_5}{2}
= \left(5\ri_2 -4 \ri_4\right) \left(\ri_2+\ri_4\right)\,,
\ee
where the label $s$ stays for `small' and the label $b$ for `big', the volume (\ref{overallVol}) simplifies to:
\be
\vo=\frac 13 \left( \sqrt{\tau_s}\, \tau_b - \,\tau_7^{3/2}\right)\,.
\label{overallVo}
\ee
We shall now show that two of the three remaining moduli can be fixed within the K\"ahler cone,
at large volume and weak coupling by the interplay of $\alpha'$
and non-perturbative corrections without fine tuning the background fluxes, i.e.\ for $W_0\simeq \mc{O}(1)$.
We shall then lift the remaining flat direction via string loop corrections to the K\"ahler potential.

Writing the K\"ahler moduli as $T_i=\tau_i + i \zeta_i$, the K\"ahler potential and superpotential
of the 4D $\mc{N}=1$ supergravity read (for $N_{gc}=3$):
\be
K=-2\ln\left(\vo+\frac{\hat\xi}{2}\right)\,,\qquad W=W_0 + A \,e^{-\frac{2\pi T_7}{N_{gc}+1}}=W_0+ A \,e^{-\frac{\pi T_7}{2}}\,.
\ee
The F-term scalar potential is given by a complicated expression similar to (\ref{PotN})
which now can very well be approximated by an expression like (\ref{PotApprN}) with slightly different
coefficients due to the fact that here we are considering $N_{gc}=3$ instead of $N_{gc}=4$:
\be
V\simeq 2\pi ^2 A^2 \frac{\sqrt{\tau_7}}{\vo}\, e^{-\pi  \tau_7}
- 2\,\pi A  W_0 \,\frac{\tau_7}{\vo^2}\,e^{-\frac{\pi  \tau_7}{2}}+\frac{3 W_0^2 \hat\xi }{4 \vo^3}\,.
\label{VAppr}
\ee
As we have already pointed out, the presence of just one D-term condition
will allow us to fix one direction exponentially large.
In this limit we can therefore safely neglect corrections proportional to $\tau_7^{3/2}/(2\vo)$ and $\hat\xi/\vo$.

The minimisation conditions in the limit $\pi\tau_7/2\gg 1$
give the following solution:
\be
\vo\simeq\frac{ W_0 \,\sqrt{\tau_7}}{2 \pi  A}\,e^{\frac{\pi  \tau_7}{2}}\,,\qquad \text{and}\qquad
\tau_7 \simeq
\left(\frac{3 \xi}{2}\right)^{2/3} \frac{1}{g_s}\,.
\label{Vfinal}
\ee
We realise that only the particular combination of $\tau_1$ and $\tau_5$ corresponding to the
overall volume gets stabilised at this level of approximation leaving one flat direction.
We need therefore to include the subleading (due to the extended no-scale structure) $g_s$ corrections
to $K$ to obtain a trustable vacuum.

We stress that we fix the K\"ahler moduli without fine tuning $W_0$ small, and so we shall consider $W_0 \sim \mc{O}(1)$.
The choice of the volume is instead fixed by the requirement of getting a phenomenologically
viable gravitino mass (and the soft terms
generated via gravity mediation) of the order the TeV scale which forces us to consider $\vo \sim \mc{O}(10^{12})$.
Let us now present a choice of parameters that gives rise to a global AdS minimum
which breaks supersymmetry spontaneously and gives rise to the required phenomenological features:
\be
W_0= 1\,, \qquad A=0.1\,, \qquad g_s=0.05\,.
\ee
We find numerically that (\ref{Vfinal}) have a solution at:
\be
\langle\tau_7\rangle\simeq 16.37\,,   \qquad \langle\vo\rangle \simeq 0.94 \cdot 10^{12}\,,
\ee
which largely justifies the validity of our approximations.
Moreover, we can obtain TeV scale supersymmetry since the gravitino mass turns out to be of the order:
\be
m_{3/2}= e^{K/2} W_0 M_P = \sqrt{\frac{g_s}{8\pi}} \frac{W_0 M_P}{\vo} \simeq 113.95 \,{\rm TeV}\,,
\ee
and gravity mediation would yield \cite{SUSYbreaking}:
\be
M_{\rm soft}\simeq \frac{m_{3/2}}{\ln\left(M_P/m_{3/2}\right)}\simeq 3.7 \,{\rm TeV}\,.
\ee
The string scale is intermediate since it is of the order
\be
M_s \simeq \frac{M_P}{\sqrt{4 \pi \vo}}\simeq 7\cdot 10^{11} \,{\rm GeV}\,,
\ee
and so we do not recover the standard picture for GUT model building.

\subsubsection*{String-loop corrections}

Let us now show that the remaining flat direction can be lifted by the inclusion of $g_s$
corrections keeping $\tau_4$ small in order to reproduce the correct order of magnitude of
the visible sector gauge coupling. Given that the overall volume has already been stabilised
at leading order, the internal moduli space is
compact, implying a finite range for this remaining modulus.
Hence, we expect that any subleading correction does not
simply generate a runaway for this remaining field, but must
instead generically induce a minimum.

Let us now compute the intersections between
the different D7-brane stacks and the O7-plane which are crucial to determine the
form of the string loop corrections to $K$:
\bea
&&D7_a\cap D7_{k3} = \int_X J\wedge D_4 \wedge D_1 = 2\left(\ri_2+\ri_4\right) = 2 \sqrt{\tau_s}\,, \\
&&D7_a\cap D7_{gc} = -2 \ri_1 = 2 \sqrt{\tau_7}\,, \\
&&D7_a\cap D7_W =
4 \left[8 \ri_1 +19 \left(\ri_2+\ri_4\right)\right] = 4 \left(19\sqrt{\tau_s}-8\sqrt{\tau_7}\right)\,, \\
&&D7_a\cap O7 =
2 \left[2 \ri_1 + 5\left(\ri_2+\ri_4\right) \right] = 2\left(5\sqrt{\tau_s} - 2 \sqrt{\tau_7}\right)\,, \\
&&D7_{k3}\cap D7_{gc} = 0\,, \\
&&D7_{k3}\cap D7_W =
4\left(9 \ri_2+\ri_4\right) = \frac{4\left(8\tau_b+41\tau_s\right)}{9\sqrt{\tau_s}}\simeq
\frac{32}{9}\frac{\tau_b }{\sqrt{\tau_s}}\,, \\
&&D7_{k3}\cap O7 =
2\left(3 \ri_2 + \ri_4\right) = \frac{2\left(2\tau_b+17\tau_s\right)}{9\sqrt{\tau_s}}\simeq
\frac 49\frac{\tau_b }{\sqrt{\tau_s}}\,, \\
&&D7_{gc}\cap D7_W = -32 \ri_1= 32 \sqrt{\tau_7}\,, \\
&&D7_{gc}\cap O7 = -4 \ri_1 = 4 \sqrt{\tau_7}\,, \\
&&D7_W\cap O7 = 8\left(8 \ri_1 + 43 \ri_2 + 28 \ri_4\right) =
\frac{8\left(5\tau_b+104\tau_s\right)}{3\sqrt{\tau_s}}-64\sqrt{\tau_7}\simeq
\frac{40}{3}\frac{\tau_b }{\sqrt{\tau_s}}\,.
\eea
Due to the fact that $D7_{k3}\cap D7_{gc}=0$, we have also KK corrections
between these two non-intersecting D7-stacks. Thus, we need
a measure for the `distance' between the brane wrapping a generic K3 fibre in the class of $D_1$ and a brane wrapping $D_7$. The K3 fibre is irreducible except for a particular point
of the $\mbb{P}^1$ base where it splits into $D_6+D_7$.
The `transverse distance' $d$ between a generic fibre and the point where it becomes reducible 
is proportional to the volume $t^{\perp}$ of the $\mathbb{P}^1$ base:
\be
  d = r \cdot t^{\perp}\,.
\ee
The factor of proportionality $r$ is the open string modulus that controls the position of the D7-brane on the $\mathbb P^1$ base. It tends to zero as the brane wrapping a K3 fibre approaches a brane wrapping $D_7$ and to one on the antipodal point on the base. This modulus is treated on the same footing as the complex structure moduli, i.e.\ it is fixed at tree level by some gauge fluxes and at this level it is a flux-dependent constant.
$t^{\perp}$ is obtained by plugging the D-term condition \eqref{Dcond2nd} into ${\rm Vol}(D_4\cap D_5) = 2(t_3-t_4)$:
\be
t^{\perp} = {\rm Vol}(\mbb{P}^1)={\rm Vol}(D_4 \cap D_5) =2 \left(\ri_2 + \ri_4\right) = 2\sqrt{\tau_s}\,.
\ee

\begin{figure}
\begin{center}
\scalebox{1.2}{\includegraphics*{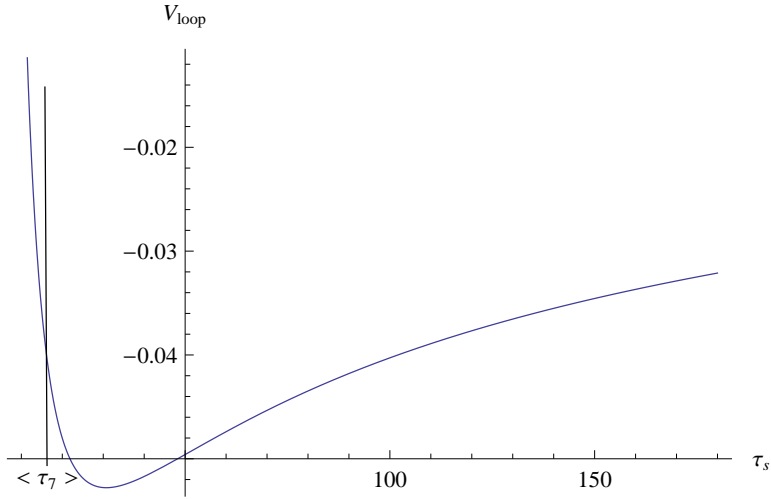}}
\caption{The potential for $\tau_s$ at fixed $\vo$, $\tau_4$ and $\tau_7$ (we rescaled $V\to \vo^3 \cdot V$). The plot
shows also how the minimum lies within the K\"ahler cone since the most constraining condition $\ti_2>0$
corresponds to $\tau_s>\tau_7$ (combined with $\sqrt{\tau_7}=- \ri_1 >0$ and $\sqrt{\tau_s}= \ri_2+\ri_4>0$
which is obtained from the condition $\ti_3>0$ after using the D-term constraint).}\label{fig2}
 \end{center}
\end{figure}

We can now use these expressions to work out the
implications of (\ref{KgsKK}) and (\ref{KgsW}) for the effective
scalar potential. The leading order result in the approximation
of large volume and small string coupling is\footnote{We neglect contributions which do not depend on the flat direction $\tau_s$
but introduce just a subleading potential for $\tau_7$ and $\vo$. The constant $c_i$ depend on the complex structure moduli and the open string moduli, that take their flus-stabilised value.}:
\be
 \delta V_{(g_s) }^{1-loop}=\left(
 \frac{c_1}{\sqrt{\tau_s}} +\frac{c_2}{5 \sqrt{\tau_s} - 2\sqrt{\tau_7}} +
 \frac{c_3}{19\sqrt{\tau_s} - 8\sqrt{\tau_7}}\right) \frac{W_0^2}{\vo^3}+\mc{O}\left(\frac{1}{\vo^4}\right),
 \label{Vloop}
\ee
where we have absorbed all the $\mc{O}(1)$ numerical factors in the parameters $c_i$, $i=1,2,3$,
which depend on the complex structure moduli. The potential (\ref{Vloop}) develops a minimum
for negative $c_1$ and positive $c_2$ and $c_3$ which is indeed located at small $\tau_s$
for natural $\mc{O}(1)$ values of the coefficients of the string loop corrections.
In fact, choosing $c_1=-1$, $c_2=2.3$ and $c_3=0.5$ we find a minimum at:
\be
\langle\tau_s\rangle \simeq 1.88 \cdot\langle\tau_7\rangle\simeq 30.74\quad \Rightarrow\quad
\langle\tau_4\rangle \simeq 4.63 \cdot\langle\tau_7\rangle \simeq 75.87\,.
\ee
The potential for $\tau_s$ at fixed $\vo$, $\tau_4$ and $\tau_7$ is plotted in Fig. \ref{fig2}.

Therefore, we managed to obtain a LVS with, at the same time,
a visible sector gauge coupling of the correct order of magnitude since
its flux corrected value turns out to be:
\be
\alpha_{\rm vis}^{-1}= \langle\tau_4\rangle+ \frac{1}{2 g_s}\int_{D_4}\mc{F}_4\wedge \mc{F}_4\simeq 136\,.
\label{avis}
\ee
We can also check that all the K\"ahler moduli are fixed within the K\"ahler cone since:
\be
r_1 = 4.05 >0\,, \qquad r_2\simeq 1.50>0\,, \qquad r_3\simeq 5.54 >0\,, \qquad r_4\simeq 2.04\cdot 10^{10} >0.
\label{KconeVEVs}
\ee
We point out that choosing the coefficients of the loop corrections in such a way
to reduce the value of $\langle\tau_4\rangle\propto \alpha_{\rm vis}^{-1}$
would also lead $r_2$ to smaller values below unity where we would loose control over our
approximations. The value of $\langle\tau_4\rangle$ could also be reduced by
lowering the VEV of $\tau_7$ choosing a larger value of $g_s$. However we would then be forced to
fine tune $W_0$ and $A$ in order to keep the same value of $\vo$ which gives rise to TeV-scale SUSY.

Moreover, the volumes of all the divisors are fixed larger than unity in a regime where we can trust
the effective field theory since\footnote{Also all the other divisors have large volumes.}:
\be
\langle\tau_1\rangle \simeq 2.23\cdot 10^{11}\,, \qquad \langle\tau_4\rangle \simeq 75.87\,,\qquad \langle\tau_5\rangle\simeq 2.26\cdot 10^{11}\,,
\qquad \langle\tau_7\rangle \simeq 16.37\,.
\ee
The values of the stabilised dual two-cycle volumes are also larger than unity which is a necessary condition
to trust the perturbative $g_s$ expansion of the K\"ahler potential:
\be
x_1\simeq 11.09\,, \qquad  x_4\simeq 2.04\cdot 10^{10}\,, \qquad  x_5\simeq 5.54\,, \qquad x_7\simeq 2.04\cdot 10^{10}\,.
\ee
Moreover, we stress that the largeness of the divisor volumes justifies the subdominant behaviour of the
string loop corrections with respect to the $\alpha'$ effects since in region around the minimum
$\delta V_{(g_s)}^{1-loop}/\delta V_{(\alpha')}\sim 0.001$.

We finally point out that the Calabi-Yau has a very anisotropic shape with two extra dimensions much larger than
the other four since at the minimum we have:
\bea
t_s\equiv t_2+t_4&=&\sqrt{\tau_s} = r_3 \simeq 5.54\,,\qquad t_b\equiv 5 t_2 - 4 t_4 = \frac{\tau_b}{\sqrt{\tau_s}}
=5r_3+9 r_4\simeq 1.84\cdot 10^{11}\,, \notag \\
&& t_1=-\sqrt{\tau_7}=-r_1\simeq 4.05\,.
\eea
Hence, we are in a situation where the overall volume is exponentially large
because, as can be seen from (\ref{KconeVEVs}), we have effectively just one two-cycle which is large $r_4\gg r_1\sim r_2\sim r_3$.
The reason why we can take $r_4$ arbitrarily large is because this two-cycle is not touched by the D-term constraint (\ref{Dcond2nd}).
The overall volume can be simplified as:
\be
\vo \simeq \frac 13\, t_b t_s^2 = \frac 13 \sqrt{\tau_s}\tau_b\,.
\ee
The anisotropic shape of the Calabi-Yau becomes manifest
if we work with the $D_3$ K3 fibre instead of $D_1$. In fact, the volume of this divisor
$\tau_3=3 t_s^2 = 3 \tau_s$ reveals that this K3 fibre is fixed small with the corresponding
$\mbb{P}^1$ base, given by $t_b$, exponentially large.

\subsubsection*{A different parameter choice with TeV-scale strings}

We could finally envisage a scenario where we slightly modify our choice of the string coupling from $g_s=0.05$
to $g_s=0.02$ (leaving $W_0= 1$ and $A=0.1$) which yields a new solution to (\ref{Vfinal}) at:
\be
\langle\tau_7\rangle\simeq 40.91\,,   \qquad \langle\vo\rangle \simeq 8.28 \cdot 10^{28}\,,
\ee
that gives rise to TeV-scale strings since:
\be
M_s \simeq \frac{M_P}{\sqrt{4 \pi \vo}}\simeq 2.35\,{\rm TeV}.
\ee
Notice that for such a large value of $\vo$, the gauge coupling of the field theory living
on $\tau_4$ becomes much larger than its previous value (\ref{avis}), producing a dangerous tension
between obtaining such a large value of $\vo$ and the correct visible gauge coupling.
However this scenario does not feature this problem since
the brane stack wrapping $D_4$ would not correspond to the visible sector. In fact,
models with TeV-scale strings require a non-supersymmetric brane configuration for the
Standard Model where supersymmetry is badly broken by construction \cite{ADDstrings}. This could be
easily achieved by adding to our set-up another `diagonal' del Pezzo divisor which
would shrink at the singularity due to the D-term condition and support a SM-like
quiver.

With this choice of parameters, we therefore managed to obtain a very anisotropic compactification similar
to the one used in \cite{ADDstrings} to derive string vacua with TeV-scale strings
and micron-sized extra dimensions.
The difference between our model and the one studied in \cite{ADDstrings}
is the use of string loop corrections instead of polyinstanton effects in order to fix the flat direction
left over after the effect of $\alpha'$ and ordinary non-perturbative corrections.
The reason why in \cite{ADDstrings} it has not been possible to obtain an anisotropic configuration
via $g_s$ effects is because the authors did not perform an explicit orientifold projection.
Hence, they neglected the string loops due to the exchange of winding strings at the
intersection between O7-planes and D7-branes.

We finally point out that a K3 fibration example, very similar to the one
described here, has been used to derive a very interesting inflationary scenario
which yields detectable gravity waves \cite{FibreInfl}.
The inflaton is the modulus whose potential is loop-generated and
in order to provide observable density fluctuations
the set of underlying parameters should be chosen so to
obtain a value of the overall volume of the order $\vo\sim\mc{O}(10^3-10^4)$
which would also give rise to GUT theories. We shall leave
the detailed study of the inflationary dynamics in our model for future investigation.

\section{Conclusions}
\label{Concl}

In this paper we outlined a general strategy to combine K\"ahler moduli
stabilisation with chiral D7-brane models within the
framework of Type IIB flux compactifications.
The powerful tools of toric geometry allowed us to present an example
of a compact K3-fibred Calabi-Yau orientifold where
we could perform some explicit choices of brane set-ups and world-volume
fluxes that gave rise to GUT- or MSSM-like theories. They
satisfied global consistency conditions like tadpole and Freed-Witten anomaly cancellation
or K-theory constraints. At the same time,
we managed to stabilise all the K\"ahler moduli
inside the K\"ahler cone. We did this within the regime of
validity of the low-energy theory without
facing any problem related either to D-term induced shrinking of
some divisors or to the presence of chirality
or to the cancellation of Freed-Witten anomalies.
Moreover, the VEVs of the K\"ahler moduli are such that
we obtained three different models with interesting
phenomenological scales: the first with ordinary GUT
theories and TeV-scale SUSY; the second with
TeV-scale SUSY and an exponentially large value of the
Calabi-Yau volume; and the third with TeV-scale strings
and two micron-sized extra dimensions.

We point out that the last two models represent the
first realisation of the popular LARGE Volume Scenario
for globally consistent chiral models in explicit
compact Calabi-Yau backgrounds.

Even if there are still
several issues which should be addressed in the future, we believe that this paper represents already a big step
forward. In fact, we have built not just a scenario, but a full model
where moduli stabilisation can eventually be combined
with a fully realistic D-brane set-up.

Some of these directions for future work are: $(i)$ the explicit turning on of bulk three-form fluxes
that fix the complex structure, the dilaton and the deformation moduli, and that fulfill
the D3-tadpole cancellation condition;
$(ii)$ the realisation of the correct chiral spectrum and Yukawa couplings;
$(iii)$ gauge coupling unification for GUT theories;
$(iv)$ the derivation of a Minkowski vacuum;
$(v)$ the detailed description of the inflationary dynamics of the model with just one
D-term condition which is very similar to the scenario presented in \cite{FibreInfl};
$(vi)$ the study of the phenomenology of light hidden sector particles \cite{HiddenPhotons}.

We stress that we presented just one example in detail but
we found many more chiral global models where the K\"ahler moduli
could be fixed along the same lines described here leading to similar
phenomenological features. We believe that many models
among the ones listed in \cite{CKM} present the same characteristics as the one described here.
Hence, our mechanism to fix the moduli is indeed rather general.

We finally point out that
our internal manifold features an interesting F-theory uplift
in terms of an elliptically fibred Calabi-Yau four-fold \cite{Collinucci:2008zs,Collinucci:2009uh,Blumenhagen:2009up}.
It would also be interesting to investigate if our general mechanism
can be applied successfully to F-theory.

\section*{Acknowledgements}

We would like to thank Andreas Braun, Andres Collinucci, Fernando Quevedo, Andreas Ringwald, Nils-Ole Walliser and Timo Weigand for useful discussions.
The work of C.M. was supported by the DFG through TRR33 The Dark Universe.
The work of R.V. was supported by the German Science Foundation (DFG) under the Collaborative Research Center (SFB) 676.

\end{document}